\documentclass[twocolumn,tighten,twocolappendix,apj]{openjournal}
\usepackage{aas_macros}
\usepackage{graphicx}
\graphicspath{ {images/} }

\usepackage{dcolumn}
\usepackage{enumitem}

\usepackage[utf8]{inputenc}
\usepackage[T1]{fontenc}
\usepackage{mathptmx}
\usepackage{tabularx}
\newcolumntype{Z}{>{\centering\arraybackslash}X}
\usepackage{booktabs}
\usepackage{amsmath,mathpazo}
\usepackage{xcolor}
\usepackage{xspace}

\definecolor{xlinkcolor}{cmyk}{1,1,0,0}
\PassOptionsToPackage{hyphens}{url}
 \usepackage[bookmarks=true,         
    pdfnewwindow=true,      
    colorlinks=true,    
    linkcolor=xlinkcolor,     
    citecolor=xlinkcolor,     
    filecolor=xlinkcolor,  
    urlcolor=xlinkcolor,      
    final=true,
]{hyperref}

\definecolor{diffmah_green}{RGB}{39, 217, 39}
\definecolor{diffmah_red}{RGB}{240, 23, 0}
\definecolor{diffmah_blue}{RGB}{30, 140, 255}

\newcommand{\diffmah}[0]{\textsc{diffmah}\xspace}
\newcommand{\smacc}[0]{\textsc{smacc}\xspace}
\newcommand{\hMsun}[0]{$\smash{h^{-1}}\mathrm{M}_\odot$\xspace}
\newcommand{\Msun}[0]{$\mathrm{M}_\odot$\xspace}
\newcommand{\hMpc}[0]{$\smash{h^{-1}}\mathrm{Mpc}$\xspace}

\shorttitle{Cosmo-Paleontology}
\shortauthors{Cossairt et al.}
\journalinfo{}

\begin{document}
\title{Cosmo-Paleontology: Statistics of Fossil Groups in a Gravity-Only Simulation}
\author{Aurora Cossairt$^{1,\star}$}
\author{Michael Buehlmann$^{1}$}
\author{Eve Kovacs$^{1}$}
\author{Xin Liu$^{1,2}$}
\author{Salman Habib$^{1,3}$}
\author{Katrin Heitmann$^{1}$}

\affiliation{$^1$High Energy Physics Division, Argonne National Laboratory, Lemont, IL 60439, USA}
\affiliation{$^2$Department of Physics, University of Chicago, Chicago, IL 60637, USA}
\affiliation{$^3$Computational Science Division, Argonne National Laboratory, Lemont, IL 60439, USA}
\thanks{$^{\star}$Corresponding author: acossairt@anl.gov}

\begin{abstract}
We present a detailed study of fossil group candidates identified in ``Last Journey'', a gravity-only cosmological simulation covering a $\smash{(3.4\, h^{-1}\mathrm{Gpc})^3}$ volume with a particle mass resolution of $\smash{m_p \approx 2.7 \times 10^9}$\,\hMsun. The simulation allows us to simultaneously capture a large number of group-scale halos and to resolve their internal structure. Historically, fossil groups have been characterized by high X-ray brightness and a large luminosity gap between the brightest and second brightest galaxy in the group. In order to identify candidate halos that host fossil groups, we use halo merger tree information to introduce two parameters: a luminous merger mass threshold ($M_\mathrm{LM}$) and a last luminous merger redshift cut-off ($z_\mathrm{LLM}$). The final parameter choices are informed by observational data and allow us to identify a plausible fossil group sample from the simulation. The candidate halos are characterized by reduced substructure and are therefore less likely to host bright galaxies beyond the brightest central galaxy. We carry out detailed studies of this sample, including analysis of halo properties and clustering. We find that our simple assumptions lead to fossil group candidates that form early, have higher concentrations, and are more relaxed compared to other halos in the same mass range. 
\end{abstract}


\section{Introduction}\label{sec:intro}

The formation of structure over cosmic time provides a key source of information to infer and constrain cosmological models. Due to nonlinear gravitational effects that become increasingly important as density perturbations grow, large N-body simulations are necessary for understanding the growth of structure over a wide range of scales. Analyzing observations using models based on the results of numerical simulations has emerged as a powerful phenomenological method for constraining the underlying cosmology, using a variety of observational probes (recent examples include \citealt{kids2021, des2021cosmology}).

A great deal of progress in constraining cosmology has been made by studying the statistical distributions of tracers of the matter density field (as well as the field itself), primarily using two-point statistics (and abundance in the case of clusters). The large volume of modern surveys has reduced the statistical error in many cases to the percent level, providing sufficiently robust information to simultaneously distinguish and constrain various cosmological models. At the same time, the observational information is sufficiently rich that we can also use more targeted studies of sub-populations of objects to test the underlying paradigms of the structure formation model, albeit in more qualitative ways. One such class of objects is comprised of fossil groups, which have been studied both observationally and in simulations over the last three decades \citep[studies include e.g.][]{Ponman1994, Vikhlinin1999, Jones2000, Jones2003, DOnghia2005, DOnghia2007, vonBendaBeckmann2008}; for a recent review on FGs, see \citet{Aguerri2021}.

The term \emph{fossil group} (FG) was first proposed by \citet{Ponman1994} to describe an apparently isolated elliptical galaxy embedded in a group-scale halo. The definition was refined by \citet{Jones2003}, who formalized FGs as galaxy systems with 1) a bright, extended X-ray halo of luminosity $L_{X,bol} \geq 0.25 \times 10^{42} \; h^{-2}\mathrm{ergs}^{-1}$ and 2) an absolute difference in R-band magnitude between the first and second brightest galaxies of $\Delta m_\mathrm{12}\geq2$. The X-ray criterion ensures that halos are group-sized or larger, excluding normal isolated elliptical galaxies, while the magnitude gap criterion ensures the systems are dominated by their brightest galaxies. Defined in this way, FGs exhibit some striking features. They have a mass-luminosity ratio greater than that of most galaxy clusters and their brightest galaxies have similar masses and X-ray properties to the brightest cluster galaxies, yet the number of galaxies in these systems classifies them as groups \citep{Ponman1994, Vikhlinin1999}. Further, although FGs are characterized by a very bright central elliptical galaxy \citep{DOnghia2005}, their luminosity functions show a lack of L* galaxies \citep{DOnghia2007}. Since \citet{Ponman1994}'s original discovery, much work has been dedicated to understanding the cause of these characteristics, and how to best understand FGs within the standard theory of cosmological structure formation.

Historically, FGs have been interpreted in various ways, including: 1) that they are not a special class of halos, but rather represent the final stage of mass assembly for all group scale halos \citep[e.g.][]{LaBarbera2009, Ponman1994}; 2) that they are distinct in constituting the ``oldest and most undisturbed galaxy systems not yet absorbed by larger halos'' ( \citet{Santos2007}, see also \citet{Ponman1994}); and 3) that FGs are simply ``extreme systems produced following the current theory of structure formation in the universe'' \citep{Aguerri2021}. Although FGs have been studied for decades, there is still not a firm consensus regarding the place these objects hold in the wider picture of halo formation.

Different scenarios have also been proposed to describe the formation of FGs themselves. In the \emph{merging scenario}, they are interpreted to be the relics of galaxy mergers with the following specific properties: 1) $\geq 4$\,Gyr since the last major merger, and  2) all L* satellites have fallen into the central galaxy and have been cannibalized, forming a single extremely bright elliptical \citep{Jones2000}.  Other suggestions include the \emph{compact-group} scenario \citep{Barnes1989, Farhang2017} and the monolithic or \emph{failed-group} scenario \citep{Mulchaey1999}. Based on observational studies of the brightest FG central galaxies -- including properties such as their metallicity gradients \citep{Eigenthaler2013}, location in the fundamental plane of ellipticals, and lack of strong Balmer absorption lines \citep{Jones2000} -- the current consensus strongly favors the merging scenario.

Simulations have provided further insight into various aspects of FG formation. In particular, gravity-only simulations enable statistically meaningful studies using large simulation volumes at sufficiently high mass resolution. These simulations provide the full mass accretion history of halos as well as their properties (e.g. concentration, relaxation state) at each step of the evolution. The results of gravity-only simulations in connection with hydrodynamical simulations \citep[e.g.][]{DOnghia2005} and semi-analytic models \citep[e.g.][]{Dariush2010} have shown that FGs obtain a large fraction of their final mass already at high redshifts, and that their magnitude gaps are at least weakly correlated with early formation time. A study by \citet{vonBendaBeckmann2008} arrived at a similar conclusion using halo circular velocities to predict magnitude gaps for a gravity-only simulation.

These results suggest a mechanism that is consistent with the merging scenario, in which FGs form early, giving their central galaxies sufficient time to cannibalize in-fallen satellites and create the characteristic magnitude gap. However, \citet{Dariush2010} found that most early-forming halos do not have a large enough gap to be considered as FGs. Thus, early formation is unlikely to be the sole indicator for the special properties of fossil groups. Conversely, a large luminosity gap does not guarantee that a halo formed early \citep[as shown by, e.g.][]{Dariush2010, Raouf2014}, so the \citet{Jones2003} criteria for finding FGs does not always yield an old sample of halos. The lack of a large, homogeneous, statistically clean and well-defined population of FGs remains a significant observational problem \citep{Aguerri2021} that prevents the formulation of definitive conclusions regarding the formation mechanism.

The dynamical properties of halos may not be the only drivers for FG formation. It has also been suggested \citep{DOnghia2005, Ponman1994} that the local cosmic environment may play a role -- FGs being more likely to form in an isolated environment containing fewer objects with which to merge. Here too, results from cosmological simulations have yet to provide a resolution. \citet{vonBendaBeckmann2008} found no strong environmental preferences among their FG sample, suggesting that previous observational results suffered from selection bias. Meanwhile, using the Millennium Simulation, \citet{DiazGimenez2011} found that at $z \geq 3.6$ FGs actually live in denser environments than non-fossil systems, but that this trend reverses by $z=0$.

In this paper, we investigate fossil groups by using results from a very large, state-of-the-art gravity-only simulation, the Last Journey run \citep{Heitmann2020}, combined with expectations from galaxy modeling based on the galaxy-halo connection (for a recent review, see \citealt{2018ARA&A..56..435W}). 
The simulation spans a (3.4 $h^{-1}$Gpc)$^3$ volume with a mass resolution of $m_p \approx 2.7 \times 10^9$\,\hMsun. These specifications yield excellent statistics for group and cluster-sized halos as well as the ability to track halo substructure, thus offering a solid foundation for the study of FGs. 

The major drawback of any gravity-only simulation is the lack of direct modeling of the galaxy population. For this reason, we do not employ a traditional magnitude gap criterion (i.e. $\Delta m_\mathrm{12}$ as in \citet{Jones2003}) to identify FG candidates. Instead, we use simple galaxy-halo connection ideas to model a scenario of FG formation based on hierarchical structure formation and search for candidates using halo merger trees. Models describing the galaxy-halo connection have improved continuously with the increased resolution available in contemporary simulations and may be confidently leveraged for FG studies. This approach is in keeping with the spirit -- if not the letter -- of the \citet{Jones2003} definition, which was intended to identify objects for which bright satellite galaxies had largely merged into the central dominant galaxy. Our method also avoids the complications that arise from predicting galaxy luminosity from proxies (e.g. halo circular velocity) or other parameterized methods (e.g. semi-analytic modeling, subhalo abundance matching).

We develop plausible selection criteria, based on assumptions derived from the galaxy-halo connection for FG formation, to identify a set of viable FG candidates. Since FGs are characterized by a dearth of L* galaxies, we require that FG candidates have experienced no recent major mergers by imposing a redshift cut on major merging activity. We then select a halo mass threshold to define a major merger. This mass threshold corresponds to a limit below which we would not expect a halo to host an L* galaxy. With these criteria in hand, we assemble samples of FG and non-FG candidate halos and compare and contrast their properties, merger and mass accretion histories, dynamical states, environments, and clustering behavior. This analysis allows us to investigate whether the putative FG host halos we identified have a special place among other more generic halos.

This paper is organized as follows. First, in \autoref{sec:last_journey}, we provide a brief overview of the cosmological simulation and relevant outputs underlying our studies. In \autoref{sec:selection} we describe the criteria used to identify FGs in our simulation and compare our abundances with observational estimates. We then present a range of results comparing our FG sample to non-FG halos in \autoref{sec:results}. We include findings about the merging and formation histories, substructure content, the cosmic environment in which FGs reside, and the clustering behavior of FGs. We conclude with a discussion of our results in \autoref{sec:conclusion} and outline possible further steps in the investigation of FGs.

\section{Simulation and Data Products}
\label{sec:last_journey}

\subsection{Last Journey Simulation}

Our study is based on the Last Journey simulation \citep{Heitmann2020}, which was the last full-machine run on the Mira supercomputer at the Argonne Leadership Computing Facility prior to its retirement.  Last Journey evolves $\sim 1.24$ trillion particles in a $(3.4 \, h^{-1}\mathrm{Gpc})^3$ volume, leading to a mass resolution of $m_p \approx 2.7 \times 10^9$\,\hMsun; group-scale halos are therefore comfortably resolved with many thousands of particles. The simulation was run with the N-body Hardware/Hybrid Accelerated Cosmology Code \citep[HACC,][]{Habib2016}. The simulation particles are initialized at redshift $z=200$ using the Zel’dovich approximation \citep{Zeldovich1970}, and then evolved to the present day ($z=0$) via a hybrid method using spectral particle-mesh (long-range forces) and tree (short-range forces) algorithms, combined with symplectic time-stepping. 

We assume a flat $\Lambda$CDM model with massless neutrinos and parameters consistent with recent Planck measurements \citep{Planck2020}: 
\begin{align}
\begin{split}
 \Omega_\mathrm{m} &= 0.31, \\
 \Omega_\mathrm{b} &= 0.04933, \\
 \Omega_\Lambda &= 0.69, \\
 H_0 &= 67.66 \; \mathrm{km} \, \mathrm{s}^{-1} \, \mathrm{Mpc}^{-1}, \\
 \sigma_8 &= 0.8102, \\
 n_s &= 0.9665, \\
\end{split}
\label{eq:cosmological_params}
\end{align}
where $\Omega_\mathrm{m}$, $\Omega_\mathrm{b}$, and $\Omega_\Lambda$ are the density parameters of the total matter (including baryon and cold dark matter components), baryonic matter, and dark energy content of the Universe today; $H_0$ is the Hubble constant; $\sigma_8$ specifies the amplitude of density fluctuations; and $n_s$ is the (fixed) spectral index of the primordial power spectrum.

Once the simulation reaches $z=10$, we identify halos using a friends-of-friends (FoF) algorithm with the linking length parameter $b=0.168$ for 101 time steps evenly spaced in $\log(a)$ between $z=10$ and the present epoch ($z=0$). We store information about halo properties -- including positions, velocities and masses -- starting with halos that have at least 20 particles. For each halo with at least 80 particles, we identify the 50 particles closest to its center and follow the evolution of these ``core'' particles throughout the rest of the simulation. This approach provides information about substructure evolution in halos. The Last Journey simulation has $\sim 10^6$ halos above a halo $z=0$ mass of $10^{14}$\,\hMsun and $\sim 15.8 \times10^6$ halos with $z=0$ masses in the range $[10^{13}$,\,$ 10^{14}]$\,\hMsun. The numbers of halos in specific mass bins are given in \autoref{tab:fossil_groups_found}. Once a core falls into another halo, we continue to track it as substructure. The approach is described in detail in \citet{Sultan2021} where we introduced \smacc, the Subhalo Mass-loss Analysis using Core Catalogs. \smacc incorporates the \citet{vandenBosch2005} mass loss model, given by:
\begin{equation}
m(t + \Delta t) = \begin{cases}
m(t)\exp \left[-\Delta t/\tau \right] & \zeta=0\\
m(t) \left[ 1 + \zeta(\frac{m}{M})^\zeta \frac{\Delta t}{\tau} \right]^{-1/\zeta} &\text{otherwise}
\end{cases}
\label{eq:SMACC}
\end{equation}
where $M=M(z)$ and $m=m(z)$ are the parent halo mass and subhalo mass at redshift $z$ respectively, $\tau = \tau(z)$ is the characteristic timescale of subhalo mass loss, defined as $\tau(z) = \tau_{\mathrm{dyn}}/\mathcal{A}$, $\tau_{\mathrm{dyn}}(z)$ is the dynamical time of the halo, and $\mathcal{A}$ and $\zeta$ are free parameters. For the mass range $[10^{13}$,\,$ 10^{14}]$\,\hMsun, $\mathcal{A} = 1.1$ and $\zeta = 0.1$ provide a good fit to masses obtained by using a subhalo finder \citep{Sultan2021}.

With substructure information provided via the \smacc approach, it is possible to measure the fraction of total mass attributable to the largest substructure,
\begin{align}
 f_\mathrm{sub,max} &= \frac{\max_i M_{\mathrm{sub},i}}{M_\mathrm{halo}}, \label{eq:fsubmax}
\end{align}
and the fraction of total mass attributable to the sum of all substructures,
\begin{align}
 f_\mathrm{sub,tot} &= \frac{\sum_i{ M_{\mathrm{sub},i}}}{M_\mathrm{halo}}.
\label{eq:fsubtot}
\end{align}

For each FoF halo, we also determine its spherical overdensity (SO) properties. We start from the FoF potential minimum center and measure overdensities in spherical shells around it.  We determine $r_{200c}$ (SO radius corresponding to a density contrast of 200 with regard to the critical density $\rho_c$), the NFW concentration $c_{200c}$, and the NFW scale radius, $r_s$. The latter two quantities arise from a parameterization of the halo radial density distribution by a Navarro-Frenk-White (NFW) profile \citep{Navarro1996, Navarro1997} of the form
\begin{equation}
    \rho(r) \propto \frac{1}{(r/r_s)(1 + r/r_s)^2},
    \label{eq:nfw_profile}
\end{equation}
where the NFW scale radius $r_s$ is defined as the radius where $\rho \propto r^{-2}$. We then define the halo concentration as 
\begin{equation}
    c_{200} = r_{200c} / r_s.
    \label{eq:concentration}
\end{equation}
Additionally, we estimate the relaxational state of a halo by measuring the relative offset distance $\Delta x$ between the potential minimum and the FoF center of mass, normalized by $r_{200c}$:
\begin{equation}
    x_\mathrm{off} = \Delta x / r_{200c}.
    \label{eq:relaxation}
\end{equation}
This relative offset distance provides a rough measure of how well the halo is virialized \citep{child2018}. Relaxed halos have, on average, small offsets between their potential minima and mass centers, while unrelaxed halos have larger offsets.

\subsection{Merger Trees}

An important aspect of our investigation is the tracking of the evolution of halos and their substructure over time. Having identified halos and their constituent particles, we construct merger trees by matching halo particle identification tags across time steps, thus tracing each halo's trajectory, changes in mass, and merger activity. Our specific merger tree implementation, which runs backwards in time in post-processing mode, is described in \citet{Rangel2020}. The information obtained is then stored in merger tree catalogs. 

\begin{figure*}[htp]
  \includegraphics[width=\textwidth]{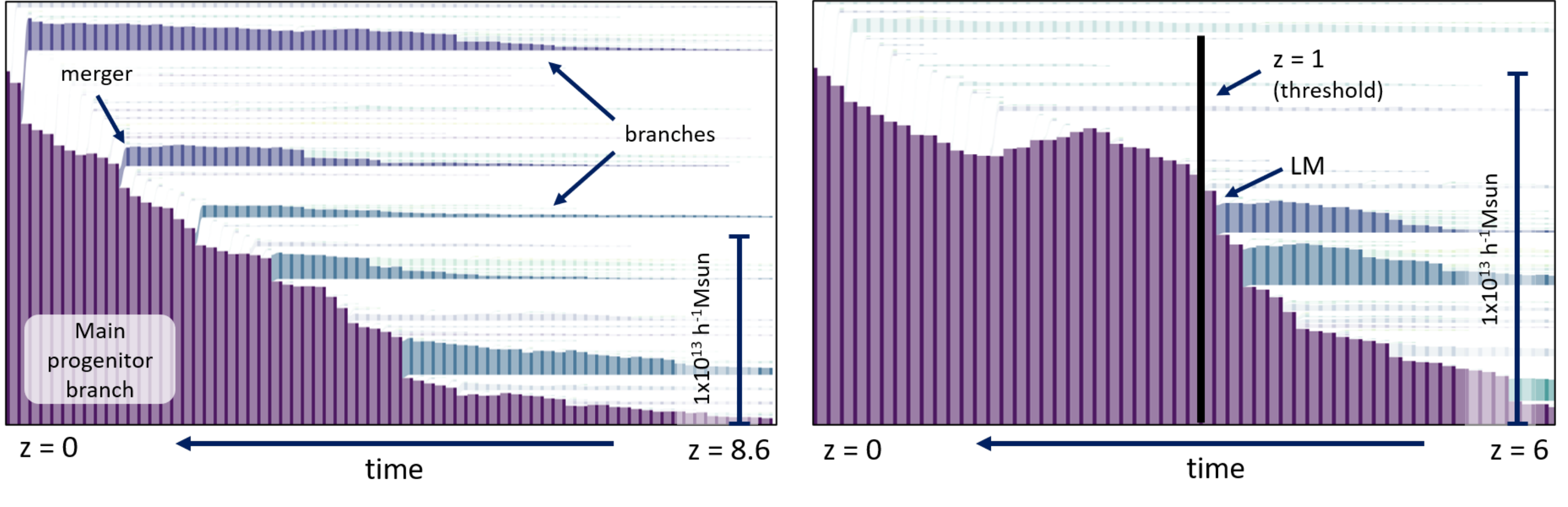}
  \caption{Mass evolution history of the main progenitor branch of a randomly selected halo with 5 luminous merger (LM) events with a final halo mass at $z=0$ of $1.98 \times 10^{13}$\,\hMsun (\textbf{left}) and an FG candidate halo with 2 luminous merger events and final mass at $z=0$ of $1.1 \times 10^{13} $\,\hMsun (\textbf{right}). Halo mass at each timestep, $M(z)$, is shown for main progenitor branch halos (dark purple) and merging halos (light purple and blue) according to the mass scale given at the right of each panel. All branches with merging mass greater (less) than our luminous merger threshold of $5 \times 10^{11}$\,\hMsun are rendered in solid (transparent) fill. In the right panel, in agreement with our definition of FG candidates, no luminous merger events occur after the redshift cutoff at $z_\mathrm{LLM}=1$, highlighted by a black vertical line -- the last luminous merger (LLM) occurs just before $z_\mathrm{LLM}$.}
  \label{fig:merger_tree_visualization}
\end{figure*}

The anatomy of an individual merger tree is illustrated in the left panel of \autoref{fig:merger_tree_visualization}. Merger trees are dominated by the main progenitor branch, which is defined by following the most massive halo progenitor at each time step. Branches -- other halo merger trees which merge with the main progenitor branch -- are shown as lighter colored limbs, and mergers are visible where smaller branches connect with the main progenitor branch. From one time step to the next, halos may increase their mass through passive accretion or mergers with other halos, or they may lose mass, e.g., due to fly-by events or particle loss on halo outskirts. We track the mass-accretion history of the main progenitor branch for each halo and identify the times at which significant merger events occurred. By varying a number of criteria, we are able to select samples of halos with specific merger-history properties in order to identify FG candidates. 

\section{Fossil Group Selection}
\label{sec:selection}

In this section we describe our approach for identifying possible FGs in our gravity-only simulation. To quantify halo merging activity, we define a \emph{luminous merger} (LM) as any merger with a halo whose mass exceeds some threshold $M_\mathrm{LM}$, above which halos are expected to contain (sufficiently) luminous galaxies (see, e.g., \citealt{yang03mnras}; \citealt{Wechsler2018}), with, say, L$\geq$L$^*$. According to the merging scenario, FGs experience a decline in merging activity at late times. We define FG candidates as those halos which experience their \emph{last luminous merger} (LLM) before a redshift cutoff $z_\mathrm{LLM}$.
The choice of $z_\mathrm{LLM}$ is somewhat arbitrary, but the cutoff certainly has to occur \emph{after} the peak of star formation activity at $z\sim 2$. This ensures that galaxies will have largely assembled their stellar mass, but leaves enough time for satellite galaxies to fall into the central galaxy, thus creating the large magnitude gap that characterizes FGs. This process of orbital decay usually occurs on a timescale of less than one Hubble time for group-mass systems \citep{DOnghia2005}.

\autoref{fig:merger_tree_visualization} illustrates this definition, comparing a random non-fossil halo to an FG candidate which experiences no luminous mergers after the redshift cutoff $z_\mathrm{LLM}=1$. Note that our definition of FG candidates includes halos whose LLM occurs before $z_\mathrm{LLM}$ as well as a subset of quiescent halo (QH) candidates that never experience a luminous merger; these objects will be discussed further below.

\subsection{Abundance of FG candidates}

In our approach, the number of halos classified as FGs depends on two parameters: $M_\mathrm{LM}$ and $z_\mathrm{LLM}$. In this section, we examine this dependence and compare our FG candidate abundances with observational results. These comparisons are necessarily approximate due to uncertainties in the assumed galaxy-halo connection, in the conversion of observational quantities such as galaxy luminosity and X-ray temperatures into halo masses via mass-observable relations, and the lack of a large controlled observational FG sample. In addition, our comparisons are carried out at $z=0$ while many observational catalogs span a range of redshifts, albeit all at $z \lesssim 0.1$.

\begin{figure}[htp]
  \includegraphics[width=\columnwidth]{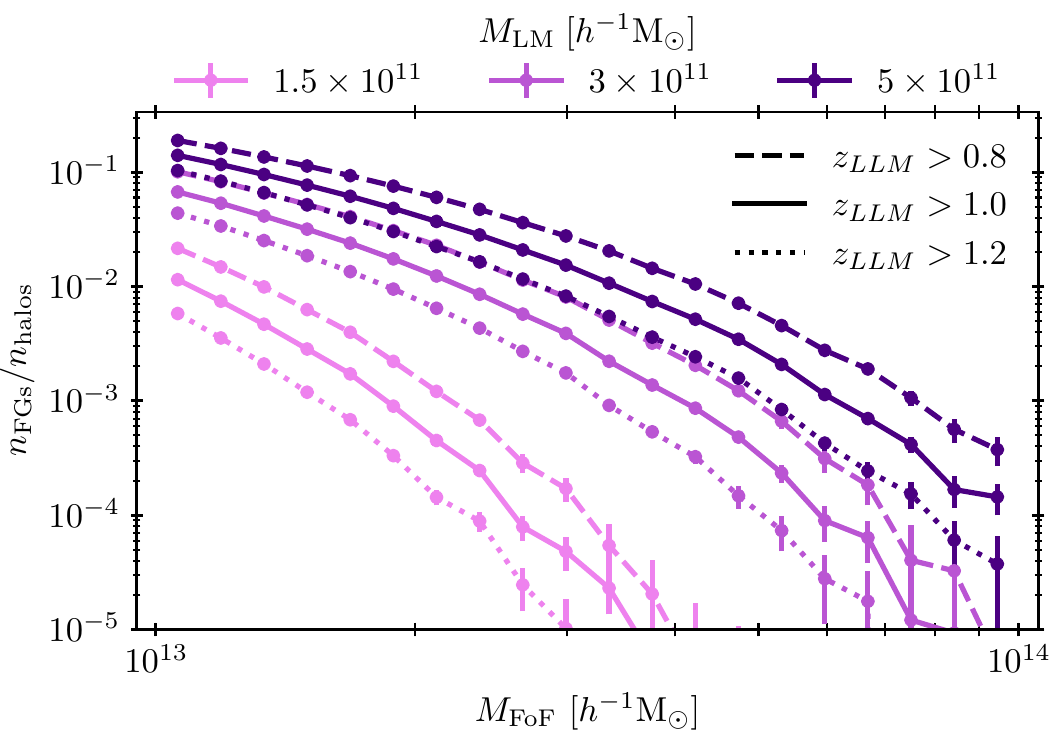}
  \caption{Fraction of FG candidates as a function of halo mass at $z = 0$ in bins of width 0.05\,dex. The fractions for luminous merger mass thresholds of $1.5 \times 10^{11}$, $3 \times 10^{11}$, and $5 \times 10^{11}$\,\hMsun are shown in light, medium and dark purple, respectively. The dashed, solid and dotted lines show the abundances obtained by setting the redshift threshold for the last luminous merger to $0.8$, $1.0$, and $1.2$, respectively.}
  \label{fig:fraction_of_candidates}
\end{figure}

\begin{table*}[htp]
 \begin{tabularx}{\linewidth}{@{}p{4cm}|XXXX|l@{}} 
  & \multicolumn{4}{c|}{groups} & clusters\\
  \toprule
  $\log_{10}(M_0 / h^{-1}\mathrm{M}_\odot)$ & $13.0$ - $14.0$ & $13.0$ - $13.05$ & $13.3$ - $13.35$  & $13.6$ - $13.65$ & $14.0+$\\
  \# particles in FoF & 3680 -- 36800 & 3680 -- 4129 & 7343 -- 8238 & 14650 -- 16438 & 36800+ \\
  \midrule
  \# halos & 15,825,664 & 1,919,589 & 975,970 & 474,295 & 1,003,955\\ 
  \# FGs & 953,215 & 269,358 & 36,181 & 2,454 & 8\\ 
  \# QHs & 60,001 & 30,077 & 294 & 0 & 0\\ 
  \midrule
  FG \% (FGs in halos) & $6.000 \pm  0.006$ & $14.03 \pm  0.03$ & $3.71 \pm  0.02$ & $0.51 \pm  0.01$ & $(8 \pm 3) \times 10^{-4}$ \\ 
  QH \% (QHs in FGs) & $6.29 \pm 0.03$ & $11.17 \pm 0.07$ & $0.81 \pm 0.05$  & 0 & 0\\
  \bottomrule
 \end{tabularx}
 \caption{Total counts of halos, FG, and QH candidates and fractions of FG and QH candidates for the fiducial parameters $M_\mathrm{LM} = 5 \times 10^{11}$\,\hMsun and $z_\mathrm{LMM} = 1$ in the range $[10^{13}, 10^{14}]$\,\hMsun and within three narrower mass bins. We also list the count for fossil clusters using the same merger history criterion. The fractions of FG  (QH) candidates are computed relative to the size of the entire sample of halos (the FG sample) in each mass bin. Errors are calculated assuming Poisson errors on the number counts. There are significantly more FG and QH candidates at lower masses compared to higher masses (cf. \autoref{fig:fraction_of_candidates}).}
 \label{tab:fossil_groups_found}
\end{table*}

\autoref{fig:fraction_of_candidates} shows the fraction of candidates at a fixed halo mass for three different choices for the mass threshold $M_\mathrm{LM}$ ($1.5 \times 10^{11}$, $3 \times 10^{11}$, and $5 \times 10^{11}$\,\hMsun) and the redshift cutoff $z_\mathrm{LLM}$ (0.8, 1.0, and 1.2).  As $M_\mathrm{LM}$ increases, so does the fraction of FG candidates -- as merger tree branches are pruned, the number of luminous mergers is reduced, and consequently the FG fraction is increased. Increasing $z_\mathrm{LLM}$ leads to a smaller fraction of FG candidates, as the number of branches pruned decreases with increasing $z_\mathrm{LLM}$.

For the rest of our analysis, we set $z_\mathrm{LLM}=1.0$, which corresponds to a lookback-time of $\sim 8$\,Gyrs (of the order of the Hubble Time) and provides enough time for relaxation of luminous substructure \citep{boylan2008dynamical, jiang2008fitting}. We select the mass threshold $M_\mathrm{LM} = 5 \times 10^{11}$\,\hMsun, because halos with this mass or larger can reasonably be expected to host an L* galaxy \citep{yang03mnras}. Additionally, the stellar-to-halo mass relation extracted from COSMOS data by \citet{girelli2020} predicts that a halo with a mass equal to $M_\mathrm{LM}$ will host a central galaxy having a stellar mass of $\sim 10^{10}$\,M$_\odot$, with a scatter of $\sim 0.2$\,dex. Galaxies with this range of stellar mass should be sufficiently bright (galaxies similar to the Milky Way have a stellar mass of $\sim 5\times 10^{10}$\,M$_\odot$) to be detectable and therefore be included in the determination of the observed magnitude gap.

In order to minimize the mass-dependence of our statistics, we examine FG candidates within three narrow mass-bins, each with a width of $0.05$\,dex. The bin edges and total counts of halos, FG candidates, and QH candidates are reported in \autoref{tab:fossil_groups_found} for each narrow mass bin, along with a wider bin encompassing $z=0$ halo masses in the range $[10^{13}$,\,$ 10^{14}]$\,\hMsun. We also report the fraction of FG candidates in the total sample and the fraction of QH candidates in the FG candidate sample. In total, we find 953,215 FG candidates in the mass range $10^{13}$ to $10^{14}$\,\hMsun, corresponding to a spatial density of $\sim 2.8 \times 10^{4}$ candidates per $(h^{-1} \mathrm{Gpc})^3$. This agrees well with \citet{Jones2003}, who estimated\footnote{We have converted the densities quoted in the following works to units of $(h^{-1}\mathrm{Gpc})^3$} $3.2 ^{+2.2}_{-1.5} \times 10^{4}$ FGs per $(h^{-1} \mathrm{Gpc})^3$ and also with various results from the Sloan Digital Sky Survey, which range from $0.4 - 2.4 \times 10^{4}$ FGs per $(h^{-1} \mathrm{Gpc})^3$ \citep{Vikhlinin1999, Santos2007, LaBarbera2009}. \citet{Adami2020} find a somewhat lower density of  $0.5\times 10^{3}$ FGs per $(h^{-1}\mathrm{Gpc})^3$. The variation in these observed spatial densities reflects the dependence of the measured abundances on the selection cuts used to define FG candidates.

As in \autoref{fig:fraction_of_candidates}, we find a much higher fraction of FG candidates in the lowest mass bin ($\sim 14$\%) compared to the highest mass bin ($\sim 0.5$\%). This low mass bin fraction is in agreement with \citet{Jones2003}, who estimated that FGs should represent 8 - 20\% of all galaxy groups of the same X-ray luminosity, using a sample of mostly low mass candidates. In the broadest mass bin, which covers all group scale masses ($[10^{13}$,\,$ 10^{14}]$\,\hMsun, we find an overall FG fraction of $\sim 6\%$ (\autoref{tab:fossil_groups_found}, column 2). This is lower than \citet{Gozaliasl2014} who found $22.2 \pm 6\%$ of their groups at $z\leq0.6$ were FGs, but agrees with \citet{vandenBosch2007} who found an FG fraction of $6.5 \pm 0.1 \%$ in the same mass bin that we use. Meanwhile, QH candidates make up about $6\%$ of all FG candidates in the sample, and this fraction decreases with increasing mass. Having established that our selection criteria leads to an FG sample that is in reasonable agreement with observational data with regard to FG counts, we will investigate the resulting sample in detail in the next section.

Owing to the excellent statistics available in our simulation, we are also able to make a prediction regarding the abundance of fossil cluster (FC) candidates, which are selected with the same criteria as FG candidates except that the masses of their host halos are required to be $\geq 10^{14}$\,\hMsun. The fraction of such objects is reported in \autoref{tab:fossil_groups_found} and corresponds to a spatial density of $\sim 0.2$ candidates per $(h^{-1} \mathrm{Gpc})^3$, making them very rare objects indeed. The small number of FC candidates that we found all have masses that lie just above the threshold for being classified as clusters, with the most massive candidate having an FoF mass of $1.6 \times 10^{14}$\,\hMsun. 

It is important to note that our use of an absolute luminous merger threshold $M_\mathrm{LM} = 5\times10^{11}$\,\hMsun to determine fossil status is highly restrictive for cluster-scale objects, and will naturally yield low abundances for FC candidates. Observational studies applying a traditional magnitude gap criterion (e.g. $\Delta m_\mathrm{12}$) have identified around two dozen FC candidates in a survey volume significantly smaller than Last Journey \citep[see e.g.][and citations therein]{Voevodkin2009, Zarattini2014, Qin2016, Pratt2016}. In comparison, we find only 8 FC candidates in the entire volume of the Last Journey simulation. While our FC candidate count is of the same order of magnitude as current observations, these statistics are likely low due to our particular fossil definition.

\section{Fossil Group Properties}\label{sec:results}

In this section, we study the properties of our FG sample in detail and present a range of results on the merger and mass evolution histories of FGs, their formation times, their concentration and relaxation distributions, and their clustering properties.

\subsection{Merger History}
\begin{figure*}[tbp]
 \includegraphics[width=\textwidth]{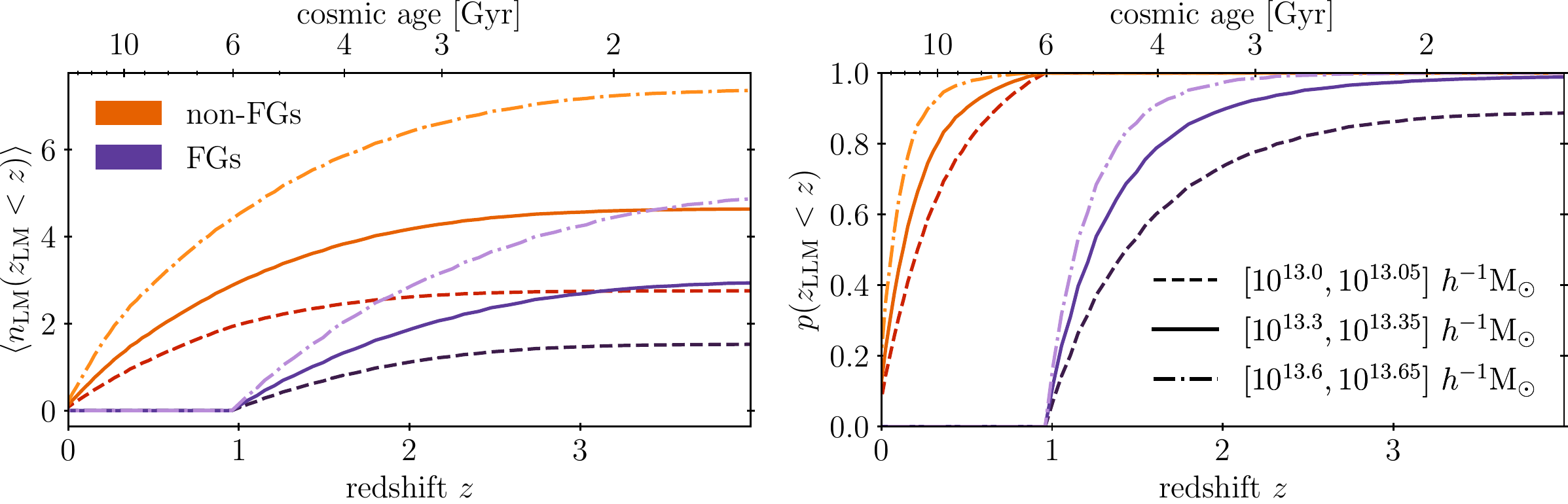}
 \caption{Mean number of luminous merger events per halo between redshift $z$ and today, $z=0$ \textbf{(left)}, alongside probability that a halo experiences its last luminous merger between redshift $z$ and today \textbf{(right)}. Two different halo populations are tested -- FG candidates (purple) and non-FGs (orange) -- across three narrow mass bins, shown by the dashed, solid, and dash-dotted lines. In agreement with our selection criterion, no luminous merging occurs for FG candidates after $z=1$. Note that QHs never experience a luminous merger and consequently some FG candidate samples have $p(z_\mathrm{LLM} < z) < 1$ even at high redshifts, in particular for the lowest mass bin (see right panel).}
 \label{fig:luminous_merging}
\end{figure*}

In \autoref{fig:luminous_merging}, we examine the redshift distributions of luminous mergers for both FG candidates and halos which do not qualify as FGs \citep[a similar measurement for major mergers defined by a relative threshold has been performed in][]{Fakhouri2010}. The left panel of \autoref{fig:luminous_merging} displays the mean number of luminous mergers $\langle n_\mathrm{LM} \rangle$ between redshift $z$ and the present day ($z=0$). The probability that a halo's \emph{last} luminous merger takes place between today and redshift $z$ is shown on the right. With our requirement that the last luminous merger occurs no later than $z=1$, FG candidates experience fewer luminous mergers cumulatively (left), and encounter their last luminous merger at a higher redshift (right) compared to non-FG candidates.  As expected, for both FG and non-FG candidates, increasing halo mass corresponds to an increase in cumulative number of mergers as well as the probability of experiencing the last luminous merger below a given redshift.

\subsection{Formation History}\label{sec:formation_history}

\subsubsection{Mass Evolution History}

Mass accretion history informs many properties of dark matter halos and the galaxies that reside within them. Halos with the same mass at $z=0$ can experience a variety of assembly histories -- a feature known as \emph{assembly bias} -- and these differences can affect their secondary properties \citep[cf. e.g.][]{Gao2005, Wechsler2006, Gao2007}.  To gain broad insight into the differences between FG candidates and non-FG-candidate halos in our simulation, we measure the mass evolution along the main progenitor branches in three narrow mass bins that have been selected to span the range of group-scale masses, but still provide excellent statistics.

\begin{figure}[htbp]
  \includegraphics[width=\columnwidth]{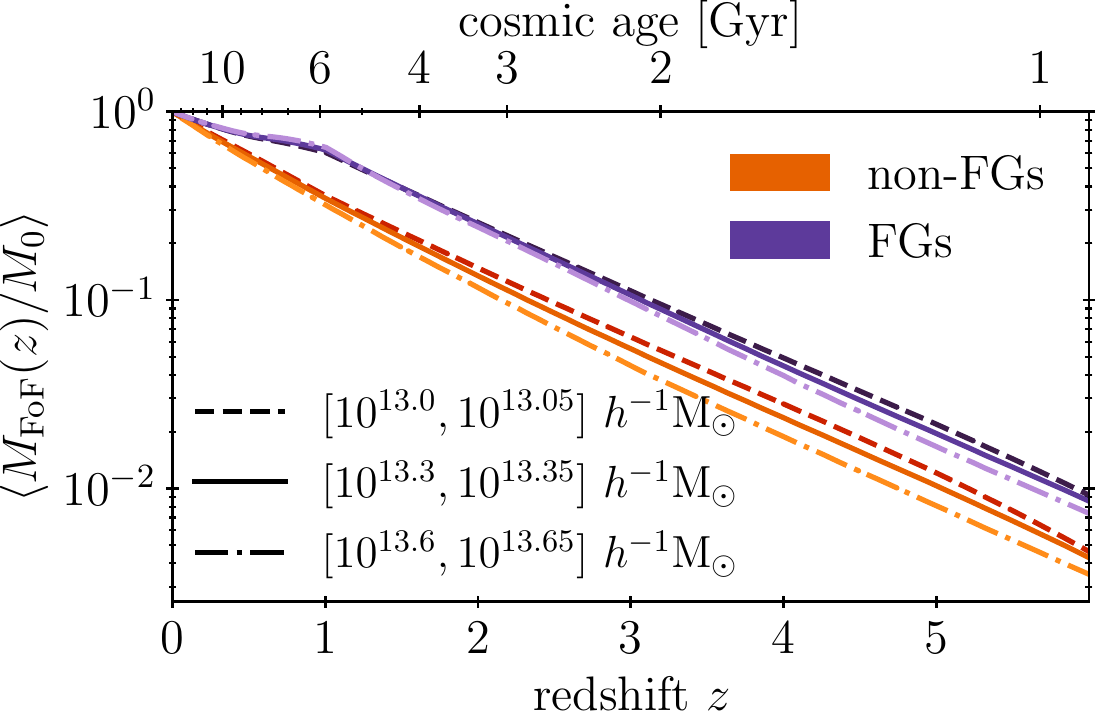}
  \caption{Mean mass evolution history $M(z)$ for halos that classify as FGs and non-FGs, averaged over narrow mass bins and normalized by present-day mass. The two samples have similar mass evolution histories for $z > 1$. For $z < 1$, the mass growth of the average FG candidate slows significantly. Unlike the non-FG sample, the mass evolution history of FG candidates is almost independent of their mass.}
  \label{fig:mass_evolution}
\end{figure}

In \autoref{fig:mass_evolution}, we compare the average mass-accretion history of FG candidates to that of non-FG candidates. At late times, due to the absence of luminous mergers, FGs grow more slowly than non-FGs of the same final mass. Because we have constrained both sample populations to the same mass at $z=0$, this late-time decline in growth results in significantly more massive FG candidate progenitors at $z>1$.  Looking back further in time, the evolution of both samples is similar, with the FG sample being offset towards higher masses. Interestingly, the shape of the mass accretion history for FG candidates is nearly independent of the final $z=0$ mass, while non-FG histories show more variation between mass bins. This behavior is likely a selection effect: by restricting luminous mergers for $z<1$, we are, in effect, requiring that FG candidates assemble most of their mass before $z=1$. This reduces the average fractional mass difference between the samples.

\subsubsection{Formation Time}
\label{sec:formation_time}

Commonly, halo age is quantified using the metric $z_\mathrm{frac}$: the redshift by which the main progenitor branch of a halo assembles some fraction of its present-day mass. For example, $z_{50}$ and $z_{80}$ denote, respectively, the redshifts by which 50\% and 80\% of the final mass of the halo has been assembled. These quantities provide simple metrics to distinguish early-forming from late-forming halos. We reiterate, however, that high values of these $z_\mathrm{frac}$ metrics, indicating early-forming halos, do not guarantee that a halo will become an FG since previous work has indicated that the magnitude gap is not formed until later times \citep{Dariush2010, Kundert2017}.

\begin{figure*}[tbp]
 \includegraphics[width=\textwidth]{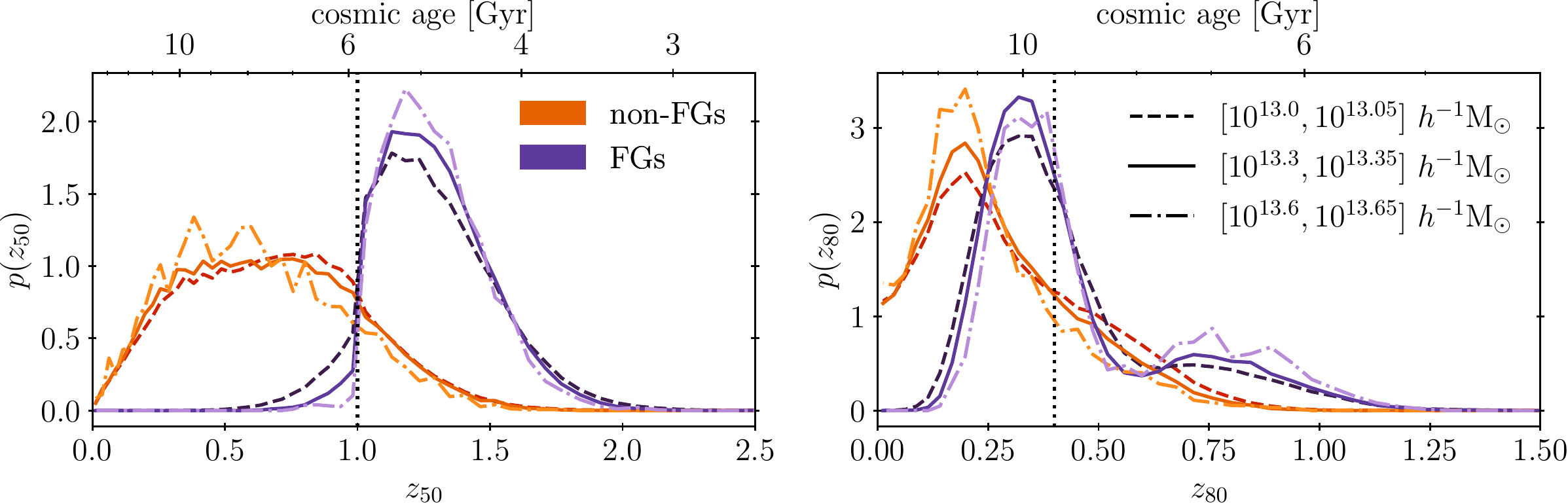}
 \caption{Probability distribution of redshifts $z$ by which halos form 50\% \textbf{(left)} and 80\% \textbf{(right)} of their present-day mass. FG candidates are compared to non-FG halos in three narrow mass bins. Vertical dotted lines provide easy comparison to the results of \citet{Kundert2017}, who found that most FGs reach 50\% (80\%) of their final mass prior to $z=1$ ($z=0.4$).}
 \label{fig:z_fracs}
\end{figure*}

In \autoref{fig:z_fracs}, we show the $z_{50}$ and $z_{80}$ probability distribution functions for FG candidates compared to non-FG halos, using the same narrow mass bins as previously defined. For both metrics, we find that FG candidates assemble a large fraction of their mass earlier than halos that do not classify as FGs.  In agreement with previous works \citep{vonBendaBeckmann2008, Zarattini2016}, we find significant differences in the $z_{50}$ distributions of FGs compared to non-FG halos, and a clear threshold in the distributions for FG candidates at $z=1$.   Almost all of our FG candidates assemble $\geq 50\%$ of their $z=0$ mass by $z=1$ (between 91\% and 99\%, depending on mass at $z=0$). However, unlike \citet{Kundert2017}, who found that the $z_{80}$ distributions exhibited stronger differences between FG candidates and non-FG halos, we find the opposite: there is less overlap of the  $z_{50}$ distributions for the two classes of objects than for the corresponding $z_{80}$ distributions. Furthermore, the redshift $z = 0.4$ identified as a threshold by \citet{Kundert2017} for the separation of the $z_{80}$ distributions, does not appear to be a significant value for distinguishing our FG candidates from non-FG halos. We note the presence of a second peak in the $z_{80}$ distribution for FG candidates above $z\sim0.6$. Comparing FGs with $z_{80} < 0.6$ to FGs with higher $z_{80}$, we find that the former group experienced their last luminous mergers at higher redshifts and contains most of the quiescent halos. FGs with $z_{80} > 0.6$ have on average a smaller separation between the potential minimum and the FoF center-of-mass, indicating a more relaxed state (cf. \autoref{sec:dynamical_state}). However, we do not find significant differences in concentration and separation between the potential minimum and the SOD center-of-mass. The reason for the ``dip'' in the $z_{80}$ distribution at $z \sim 0.6$ will be deferred to future work.

\begin{table}[htbp]
    \begin{tabularx}{\columnwidth}{@{}l|ZZZ@{}}
    \toprule
    $\log_{10} M / (h^{-1}\mathrm{M}_\odot)$ & $13.0$ - $13.05$ & $13.3$ - $13.35$  & $13.6$ - $13.65$ \\
    \midrule
    non-FGs $z_{50}$ & $0.6700 \pm 0.0006$ & $0.6501 \pm 0.0017$ & $0.6015 \pm 0.0062$ \\
    FGs     $z_{50}$ & $1.2500 \pm 0.0005$ & $1.2647 \pm 0.0011$ & $1.2629 \pm 0.0037$ \\
    \midrule
    non-FGs $z_{80}$ & $0.2849 \pm 0.0004$ & $0.2623 \pm 0.0009$ & $0.2361 \pm 0.0033$ \\
    FGs     $z_{80}$ & $0.4091 \pm 0.0004$ & $0.4208 \pm 0.0011$ & $0.4613 \pm 0.0046$ \\
    \bottomrule
    \end{tabularx}
    \caption{Mean values of $z_{50}$ and $z_{80}$ for FG candidates compared to non-FG candidates, in three narrow mass bins.} 
    \label{tab:zfrac_stats}
\end{table}

\vfil\eject\subsubsection{Differentiable Mass Accretion History (\diffmah)}
\label{sec:diffmah}

While $z_\mathrm{frac}$ offers a straightforward way to measure the formation time of halos, this metric also vastly oversimplifies the process of halo formation, which cannot be fully captured by a single statistic. In order to obtain a more complete picture of halo formation, we turn to a differentiable model of halo mass accretion history (\diffmah) \citep{Hearin2021}. This model is based on the current understanding that halo formation consists first of a \emph{fast accretion phase}, characterized by rapid mass growth and frequent merging, and then by a \emph{slow accretion phase}, in which merging is less common \citep{Bullock2001, Wechsler2002, Tasitsiomi_2004, Zhao2009}.

In \diffmah, halo growth is approximated by a power-law function of time with a rolling index,
\begin{equation}
    M_\mathrm{peak}(t) = M_0(t/t_0)^{\alpha(t)},
\end{equation}
where $M_\mathrm{peak}(t)$ is the cumulative peak halo mass at cosmic time $t$, $t_0$ is the present-day age of the universe, and $M_0 \equiv M_\mathrm{peak}(t_0)$. The use of peak mass ensures that the mass growth function increases monotonically everywhere. The power law index $\alpha(t)$ determines the rate of mass growth at time $t$, and is given by the following sigmoid function:
\begin{equation}
    \alpha(t; \tau_c, k, \alpha_\mathrm{early}, \alpha_\mathrm{late}) \equiv \alpha_\mathrm{early} + \frac{\alpha_\mathrm{late} - \alpha_\mathrm{early}}{1 + \exp\left[-k(t - \tau_c)\right]},
\end{equation}
where $\alpha_\mathrm{early}$ and $\alpha_\mathrm{late}$ are the asymptotic values of $\alpha(t)$ at early times (fast-accretion phase) and late times (slow-accretion phase) respectively. The transition between early and late regimes is governed by the transition time $\tau_c$ and transition speed $k=3.5$\,$\mathrm{Gyr}^{-1}$. In gravity-only simulations such as Last Journey, this model accurately approximates mass growth for halos of present-day mass $M_0 \gtrsim 10^{11}$\,\Msun at times $t \gtrsim 1$\,Gyr \citep{Hearin2021}.

\begin{figure*}[p]
 \centering
 \includegraphics[width=\textwidth]{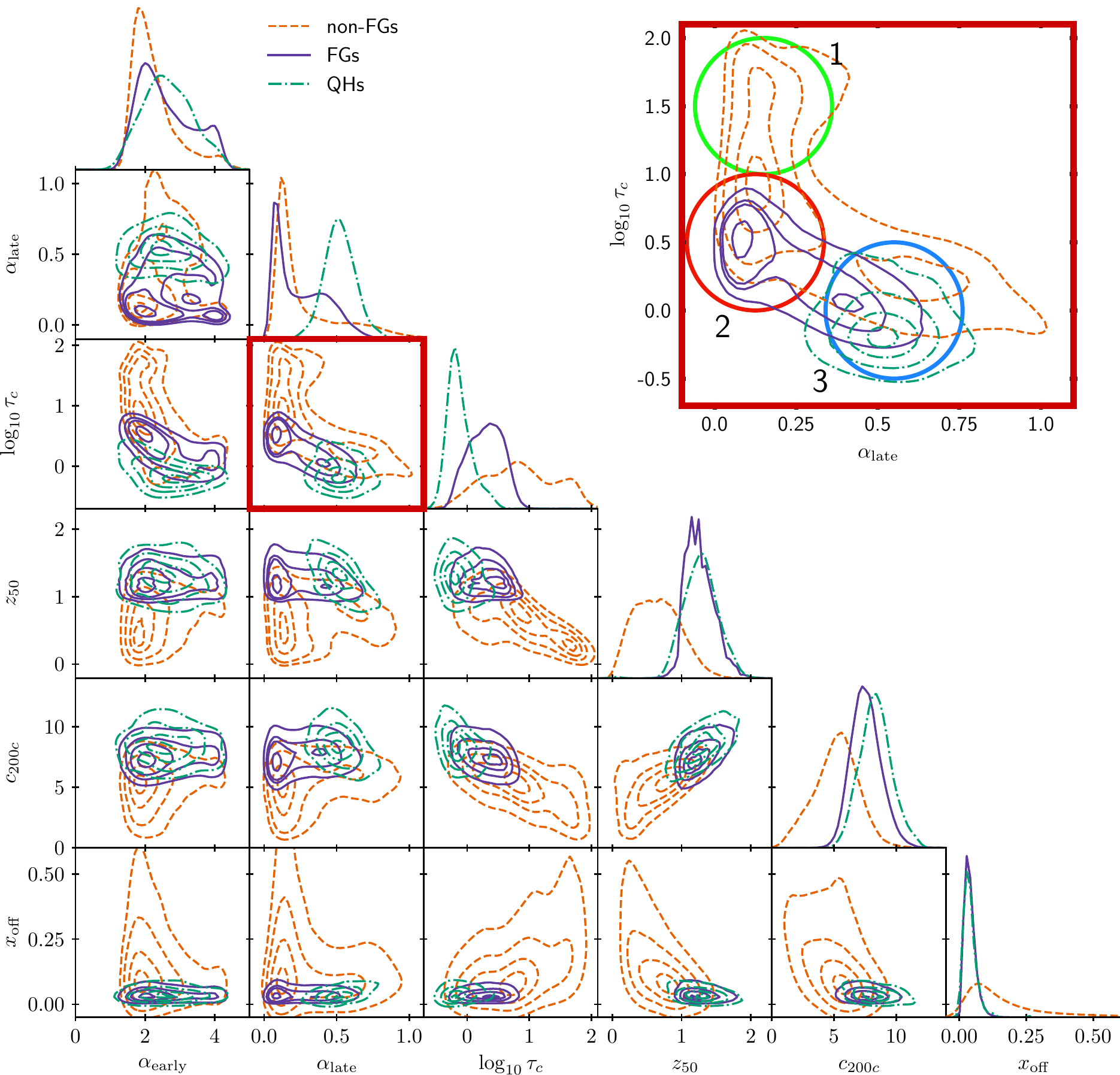}
 \caption{Two-dimensional distributions (lower triangle) and kernel density estimations (along the diagonal) of \diffmah best-fit parameters as well as $z_{50}$ and $c_{200c}$ values. The halo populations considered are mutually exclusive sets of non-candidate halos (orange, dashed), FG candidates (purple, solid), and QH candidates (green, dot-dashed). The contours enclose 90\% (outermost), 65\%, 35\%, and 10\% (innermost) of the total area of the sample in the parameter space. The inset plot in the upper right corner shows the $(\alpha_\mathrm{late}-\tau_c)$ parameter space, with three regions outlined by colored circles.  We interpret these regions as follows: 1. \textcolor{diffmah_green}{(green)} contains halos with high (late) $\tau_c$ and low $\alpha_\mathrm{late}$, so their MAHs are mostly described by their $\alpha_\mathrm{early}$ values; 2. \textcolor{diffmah_red}{(red)} contains halos with low $\tau_c$ and low $\alpha_\mathrm{late}$, so these halos accreted most of their mass very early on and are no longer experiencing much growth; 3 \textcolor{diffmah_blue}{(blue)} contains halos that transitioned from fast to slow accretion early but have high $\alpha_\mathrm{late}$, so they are still actively accreting mass today (albeit more slowly than before $\tau_c$).}
 \label{fig:diffmah}
\end{figure*}

We fit the mass accretion history with the \diffmah model for a random sample of 3000 halos evenly distributed among FG candidates (excluding QH candidates), QH candidates, and halos that do not classify as either FGs or QHs within the narrow mass bin $[10^{13.3}$,\,$ 10^{13.35}]$\,\hMsun. In \autoref{fig:diffmah}, we show the two-dimensional and one-dimensional distributions of the best-fit \diffmah parameters along with the $z_{50}$ and $c_{200c}$ distributions. We include an inset plot to interpret three key regions of the $(\alpha_\mathrm{late}, \tau_c)$ parameter space. In general, halos with smaller values of any of the three \diffmah parameters ($\alpha_\mathrm{early}$, $\alpha_\mathrm{late}$, and $\tau_c$) formed earlier than halos with larger values of the same respective parameter \citep{Hearin2021}. 

For our halo populations, \autoref{fig:diffmah} shows that FG candidates and QH candidates generally occupy similar regions of parameter space, separate from the non-FG candidates. An important exception is the case of $\alpha_\mathrm{late}$, where FG candidates and non-FG candidates display similar distributions and QH candidates diverge significantly. The parameters that exhibit the largest distinction between FG candidates and non-candidates are $\tau_c$ and $z_{50}$. For these parameters, both FG and QH candidates have much smaller values than non-FG candidates, again indicating their earlier formation times.

Some of the parameter distributions for FG candidates show a slight bimodality, which is not seen as strongly for QH and non-FG candidates. The distribution of $\alpha_\mathrm{early}$ in the FG candidates population is more evenly distributed across the interval [1, 4] than for non-FGs, with local maxima near $\alpha_\mathrm{early} \sim 2$ and $\alpha_\mathrm{early} \sim 4$. We find that FG candidates with $\alpha_\mathrm{early} > 3$ on average form later, acquire their initial mass more rapidly, and experience more luminous mergers compared to those with $\alpha_\mathrm{early} < 3$. A similar bimodal feature can be seen in the $\alpha_\mathrm{late}$ distribution as well, with local maxima near $\alpha_\mathrm{late} \sim 0.1$ and $\alpha_\mathrm{late} \sim 0.4$. We note that these features cannot be explained by QH candidates, since those were excluded from the FG candidate sample (for this figure only).

The results of these \diffmah fits provide two-dimensional distributions in a more complex mass-accretion parameter space and thus offer an in-depth view of the differences in mass accretion histories between FG and QH candidates and non-candidates. Most FG candidates are indeed early forming according to their values of $\tau_c$, but their mass accretion rates at early and late times are diverse. QH candidates, on the other hand, are almost exclusively early forming and have much higher than average growth rates at late times. It is plausible that the QH candidates -- or some subset thereof -- are hosts of isolated giant elliptical galaxies, including the isolated X-ray overluminous elliptical galaxies (IOLEGs) identified by \citet{yoshioka2003study}. In other regions of parameter space, however, the different candidate and non-candidate samples cannot be easily distinguished. Consequently, it is difficult to argue that FG and QH candidates are special, distinct classes of halos; rather they seem to lie in interesting regions of parameter space that are otherwise continuously distributed.

\subsection{Dynamical State, Substructure and Environment}
\label{sec:dynamical_state}

In this section, we compare the dynamical states, the substructure contents, and the cosmic environments of FG and non-FG candidates. We characterize the dynamical state of a halo by measuring its concentration and its degree of relaxation. The substructure is evaluated by tracking halo cores, and the environments are investigated by studying the local tidal fields surrounding each candidate.

\subsubsection{Concentration \& Relaxation}
\label{sec:relax}

\begin{figure}[htbp]
  \includegraphics[width=\columnwidth]{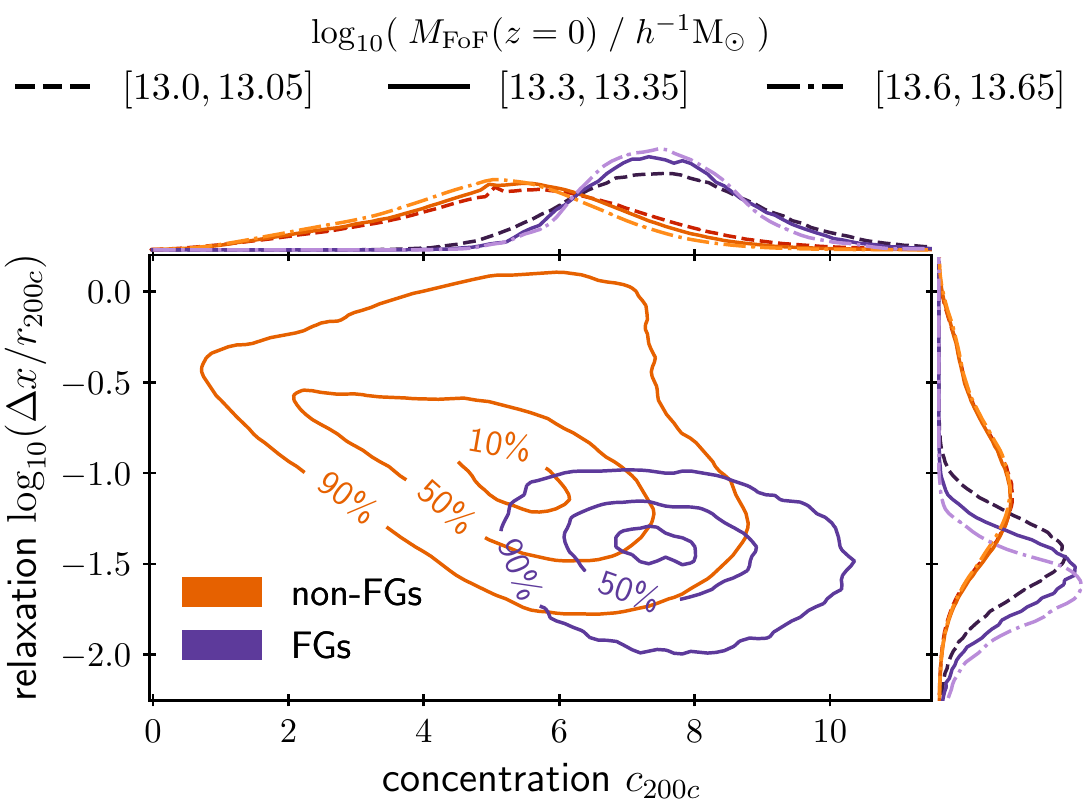}
  \caption{Two-dimensional distribution of concentrations (x-axis) and relaxations (y-axis) for FG candidates compared to non-FGs in three narrow mass bins. The contours contain 10, 50, and 90\% of the halos in the narrow mass bin $[10^{13.3}$,\,$ 10^{13.35}]$\,\hMsun, which is qualitatively representative of the other two narrow mass bins as well. One-dimensional distributions of concentration and relaxation are shown in the top and side panels, respectively, for all three narrow mass bins.}
  \label{fig:contours}
\end{figure}

Concentration parameterizes the radial mass distribution of an SO halo, giving a measure of how its mass content is peaked around its center. At the same mass, dynamically relaxed halos typically have higher concentrations than unrelaxed halos. Concentration is also correlated with the mean matter density at the time of formation: early forming objects have higher concentrations. We measure concentration $c_{200c}$ using \autoref{eq:concentration} and compare the distributions for FG and non-FG candidates for our three narrow mass bins in the top panel of \autoref{fig:contours}.

We quantify the state of relaxation of a halo by measuring $x_\mathrm{off}$ given in \autoref{eq:relaxation}. A large value for $x_\mathrm{off}$ indicates that the halo is highly asymmetric and unrelaxed,  most likely due to having experienced recent merging activity. Since we require that FGs experience no luminous mergers after $z=1$, we expect they should be more relaxed than halos that were actively merging more recently (as previously predicted by \citealt{Ponman1994} and assumed in e.g. \citealt{Khosroshahi2017}). We note, however, that the fraction of FG candidates in our sample is significantly less than the fraction of \emph{relaxed} halos in the same mass range ($\sim 75$\% for $[10^{13}$,\,$ 10^{14}]$\,\hMsun, for the large N-body cosmological simulations examined by \citealt{child2018}); the FGs essentially form a subclass of relaxed halos (Figure~\ref{fig:contours}).

In \autoref{fig:contours}, in addition to the concentration distributions mentioned above, we also show the relaxation distributions for our FG and non-FG candidates. The two-dimensional distributions are shown for one representative mass bin ($[10^{13.3}$,\,$ 10^{13.35}]$\,\hMsun), while the one-dimensional distributions are plotted for all narrow mass bins (top and side panels). We see a clear separation in both dimensions between the distributions for FG candidates and non-FGs. For the non-FGs, the average concentration decreases with halo mass, as initially observed by \citet{Navarro1996} and further quantified in the concentration-mass relation \citep[e.g.][]{child2018}. However, this is not the case with FG candidates: for these halos, the concentration distributions spread out at lower masses, but the mean value remains constant around $c_{200c} \sim 7.5$. FG candidates are also significantly more relaxed (corresponding to a lower $x_\mathrm{off}$ value) than non-FGs. According to the relaxation criterion used in \citet{Neto2007}, \citet{Duffy2008}, and \citet{child2018}, 75\%-95\% of our FGs qualify as relaxed (where the fraction of relaxed halos increases with halo mass), whereas only ~25\% of the non-FGs qualify as relaxed (independent of mass). We note that our criterion (given in \autoref{eq:relaxation}) uses the the FoF center of mass, not the SO center of mass employed by \citet{child2018}.

\begin{figure}[htbp]
  \includegraphics[width=\columnwidth]{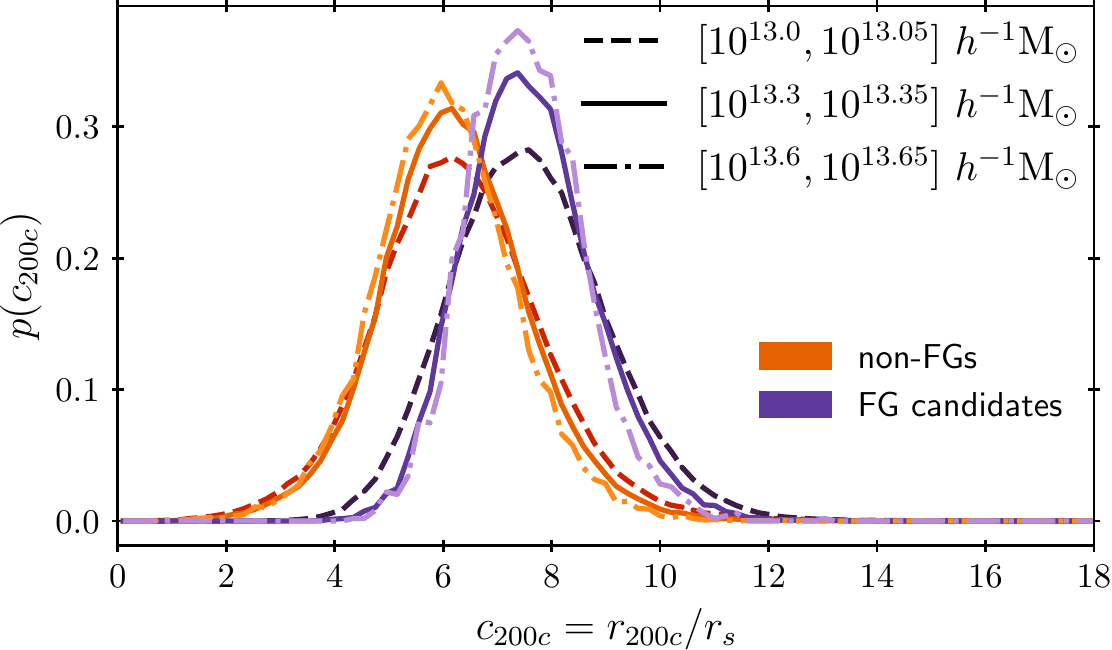}
  \caption{Distribution of concentrations of FG candidates compared to that of non-FGs across three narrow mass bins, where the non-FG sample has been matched to the FG candidates sample by relaxation distribution. We see that differences in relaxation alone cannot explain the concentration differences between the two samples.}
  \label{fig:cdeltas_corrected}
\end{figure}

Since concentration and relaxation are correlated with each other, we investigate the extent to which the differences in concentration distributions in \autoref{fig:contours} may be driven by a difference in the associated relaxation distributions between FG and non-FG candidates. In \autoref{fig:cdeltas_corrected}, we sub-sample the non-FG halos such that their relaxation distribution matches that of the FG sample. We find the resulting concentration distributions are still well separated and therefore conclude that differences in relaxation cannot independently account for differences in concentration. FG candidates thus occupy a unique region of the concentration-relaxation parameter space.

\begin{figure*}[bthp]
  \centering
  \includegraphics[width=0.95\textwidth]{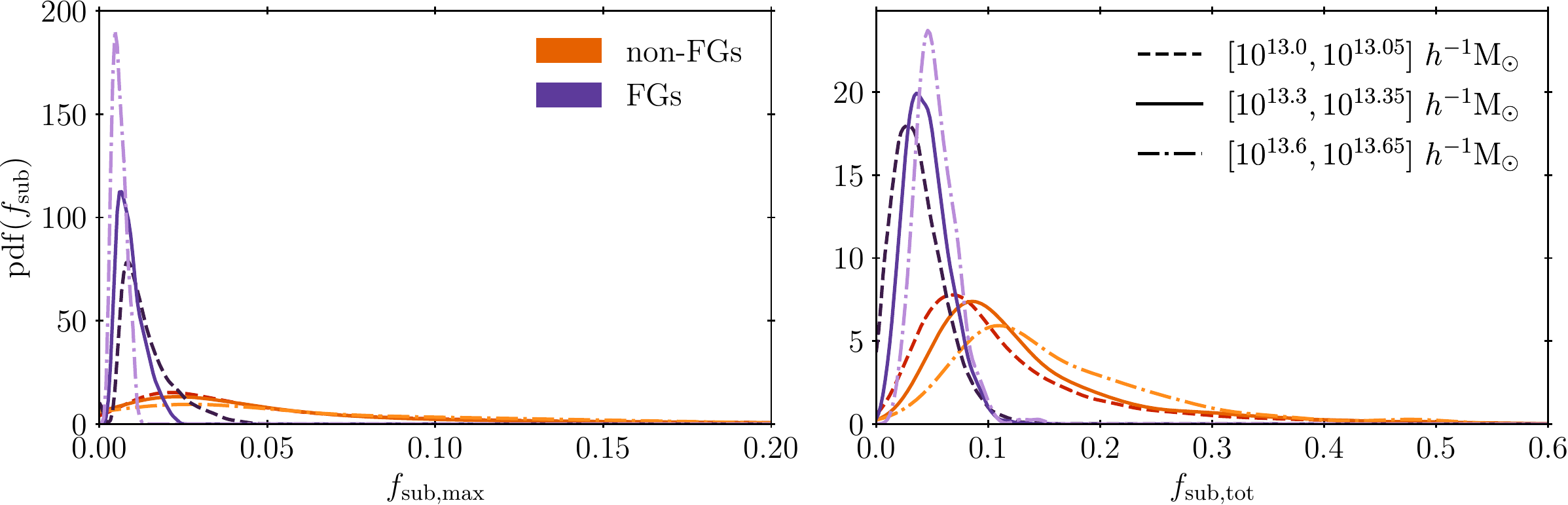}
  \caption{Distribution of the fraction of the FoF mass associated with the most massive substructure, $f_\mathrm{sub,max}$ \textbf{(left)} and the fraction of mass that is attributed to all substructures, $f_\mathrm{sub,tot}$ \textbf{(right)} (cf. \autoref{eq:fsubmax} and \ref{eq:fsubtot}). Plots are given for FG candidates (purple) and non-FGs (orange) in three narrow mass bins (see linestyles). Substructure masses are estimated by evolving the mass of merged halos using the \smacc model \citep[][see also \autoref{eq:SMACC}]{Sultan2021}.}
  \label{fig:fsub_stats}
\end{figure*}

\begin{table*}[tbh]
    \begin{tabularx}{\linewidth}{@{}l|ZZZ|ZZZ@{}} 
     & \multicolumn{3}{c|}{$10^{-3} \times \langle f_\mathrm{sub,max} \rangle$} & \multicolumn{3}{c}{$10^{-3} \times \langle f_\mathrm{sub,tot} \rangle $} \\[0.2cm]
     $\log_{10} M / (h^{-1}\mathrm{M}_\odot)$ & $13.0$ - $13.05$ & $13.3$ - $13.35$  & $13.6$ - $13.65$ & $13.0$ - $13.05$ & $13.3$ - $13.35$  & $13.6$ - $13.65$ \\
    \midrule
    FGs     & $13.91 \pm  0.05$ & $ 9.34 \pm  0.07$ & $  6.0 \pm   0.1$ & $ 38.1 \pm   0.1$ & $ 45.5 \pm   0.3$ & $ 53.5 \pm   1.2$  \\
    non-FGs & $ 64.4 \pm   0.4$ & $ 63.3 \pm   1.2$ & $ 67.3 \pm   4.5$ & $117.4 \pm   0.5$ & $136.0 \pm   1.5$ & $163.7 \pm   6.1$  \\
    \bottomrule
    \end{tabularx}
    \caption{Average fraction of mass in largest substructure $f_\mathrm{sub,max}$ and in the sum of all substructures $f_\mathrm{sub,tot}$ for FG candidates and non-FG halos in three narrow mass bins. Definitions of each statistic are given by \autoref{eq:fsubmax} and \ref{eq:fsubtot}.} 
    \label{tab:fsub_stats}
\end{table*}

\subsubsection{Substructure}

In accordance with the considerations outlined in \autoref{sec:intro},
we expect FG candidates in the Last Journey simulation to be characterized by a lack of halo substructure. 
Using the subhalo mass-evolution model presented in \citet{vandenBosch2005, Sultan2021}, we can directly estimate the mass contained in dark matter substructure (see \autoref{eq:SMACC}). We evolve the subhalo masses down to the equivalent mass of 20 particles. We then quantify the substructure distribution using $f_\mathrm{sub,max}$, the fraction of total mass attributable to the largest substructure, and $f_\mathrm{sub,tot}$, the fraction of total mass attributable to the sum of all substructures. These metrics are defined in \autoref{eq:fsubmax} and \autoref{eq:fsubtot} respectively. We expect $f_\mathrm{sub,max}$ to represent the relative mass of the second brightest group galaxy, which, on average, should be hosted in the largest halo substructure. 
Therefore, this measurement should provide insight into the potential magnitude gap: the smaller the value of $f_\mathrm{sub,max}$, the larger we expect the corresponding magnitude gap to be. Similarly, we expect that $f_\mathrm{sub,tot}$ will be much smaller for FG candidates than for non-FG candidates, indicating a lack of substructure throughout the entire FG host halo. We note that this gravity-only view of substructure distribution should only be interpreted qualitatively, since the stellar and gaseous components of a galaxy after infall will evolve differently than the surrounding subhalo (e.g. ram pressure forces, distinct initial density profiles that will be impacted differently by tidal forces).

\autoref{fig:fsub_stats} and \autoref{tab:fsub_stats} summarize the substructure statistics for FG candidates and halos that experience luminous mergers after $z=1$, in three narrow mass bins. For both metrics (\autoref{eq:fsubmax} and \autoref{eq:fsubtot}), the distributions are clearly visually distinct and the average mass-fraction in substructure is less for FG candidates by nearly an order of magnitude. In particular, FG candidates' lower values of $f_\mathrm{sub,max}$ imply that the second brightest galaxy in FG candidates is significantly smaller than in the average halo: this matches our expectations for structures with a large magnitude gap.

\begin{figure*}[p]
    \centering
    \includegraphics[width=0.83\textwidth]{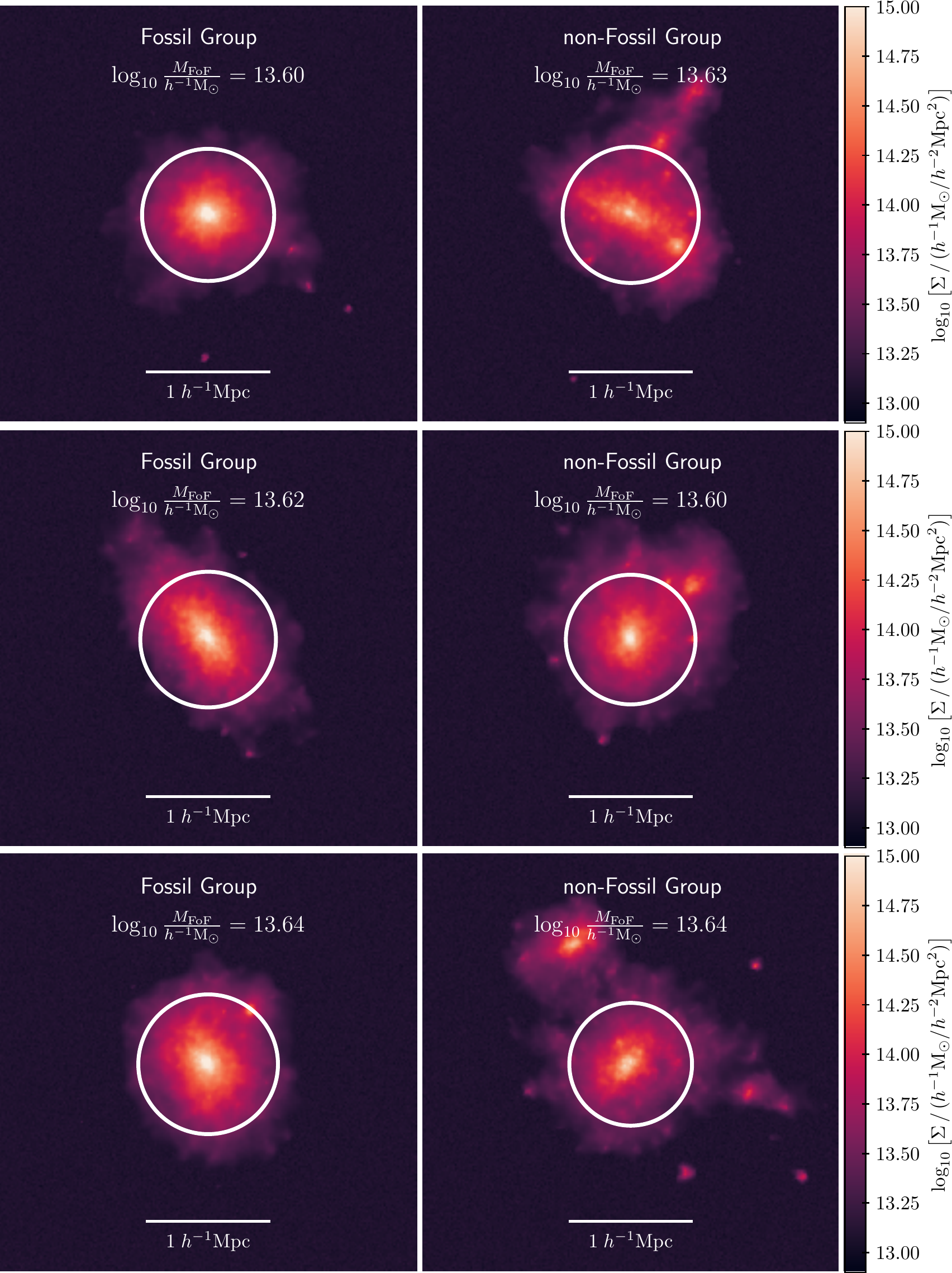}
    \caption{Surface density maps of three randomly chosen FG candidates and non-FGs from the highest mass bin. The halo radii $r_\mathrm{200c}$ are highlighted by  circles. The surface density has been estimated using the Delaunay tessellation field estimator implementation of \citet{Rangel2016}.}
    \label{fig:fg_illustrations}
\end{figure*}

To provide a visual example, we show the projected matter density distribution for three FG candidates and three non-FGs from the heaviest mass bin in \autoref{fig:fg_illustrations}. The FG candidates show significantly less substructure than the non-candidates, as expected. 

\subsection{Cosmic Environment of Fossil Groups}

In this section, we analyze multiple environmental statistics of halos -- including large-scale structure overdensity, cosmic web environment, and halo clustering -- to infer whether our sample of FG candidates demonstrates any environmental preference compared to non-FGs.

\subsubsection{Large-scale Structure Overdensity}

\begin{figure}[tbhp]
    \centering
    \includegraphics[width=\columnwidth]{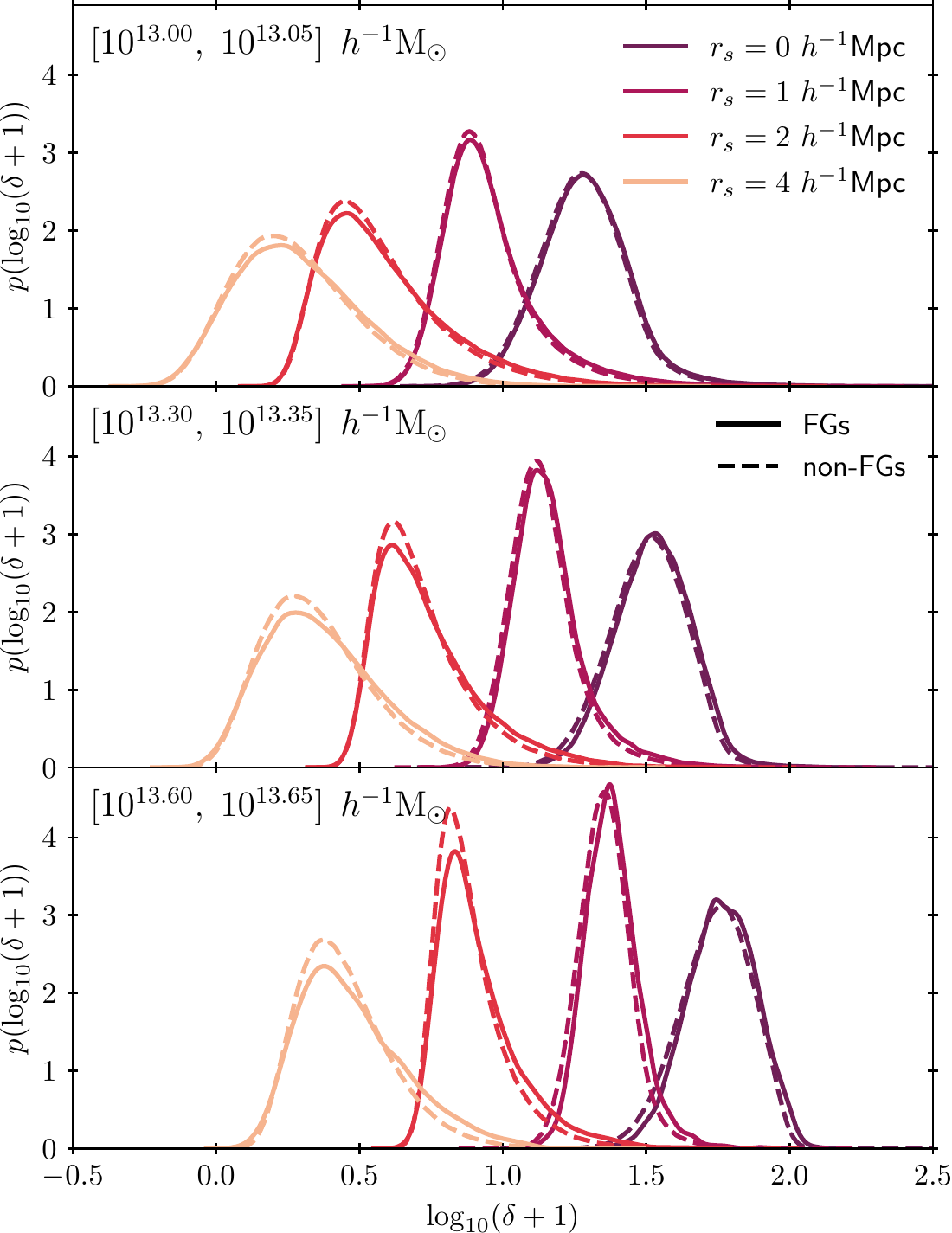}
    \caption{Distributions of the smoothed matter overdensities at the location of FG candidates (solid lines) compared to the locations of non-FGs (dashed lines). Halo samples are binned according to their $z=0$ mass (panels), and smoothed overdensities are shown for four different smoothing scales $r_s$ (colors).}
    \label{fig:smooth_overdensities}
\end{figure}

As a first environmental statistic, we measure the large-scale overdensity at the halo locations. We compute the overdensity field from the Last Journey particle data at $z=0$ on a $3072^3$ mesh using a cloud-in-cell deposition algorithm. In order to blend the halo into its environment, we smooth the overdensity field with a Gaussian kernel with radius $r_s$. This smoothing scale should be larger than the halo radius so that internal structures are removed, but not larger than the cosmic web structures in which the halo is embedded. Finding that halos in our target mass range have radii $r_\mathrm{200c} \sim 0.3 - 0.5 $\,\hMpc, we choose smoothing scales $r_s = 0, 1, 2$, and $4 $\,\hMpc. This allows us to compare the impact of $r_s$ on our results.

\autoref{fig:smooth_overdensities} shows the distributions of overdensities for FG candidates (solid) and halos that do not qualify as FGs (dashed) across different mass ranges and smoothing scales. Increasing the smoothing scale $r_s$ allows us to measure a wider environment surrounding the halo, while $r_s = 0$ measures only the local overdensity at the location of the halo itself, smoothed over the CIC cell of $\Delta x \sim 1$\,\hMpc. We see the FG candidate distribution of local overdensities is marginally shifted towards higher masses compared to non-FGs. The magnitude of these differences is negligible, although it does increase with mass and smoothing scale. This suggests that there is no significant statistical difference in the environments where FGs form compared to non-FGs, in contrast to the previous findings of e.g. \citet{Jones2003}, who found that spectroscopially confirmed FGs are found preferentially in underdense environments. This result also disagrees with \citet{DiazGimenez2011}, who found that (simulated) FG environments evolve from a state of over-density at $z\geq0.36$ to one of under-density by $z=0$. Given our much larger sample size, we agree with \citet{vonBendaBeckmann2008} that these previous observational results may have been subject to selection bias. 

\subsubsection{Cosmic Web Environment}

While the smoothed overdensity tells us about a halo's current surroundings, location in the cosmic web offers hints about its dynamical environment. As the early universe evolves, the anisotropic nature of gravitational collapse leads to the formation of a cosmic web: a network consisting of voids separated by walls, filaments, and nodes, with nodes hosting the most massive halos \citep[e.g.][]{Arnold1982b, Bond1996}. Filaments form the ``highways of the universe,'' with matter flowing along these linear structures towards nodes, which act as attractors in their environment. These dynamical arguments suggest that merging activity should be higher in nodes. However, since nodes are the peaks of the cosmic web, the halos they accrete tend to be of lower mass: this decreases the probability of luminous merging events. It is therefore unclear how a halo's proximity to a node will impact its likelihood of gaining fossil status by present day. 

A common way to classify the cosmic web in a cosmological simulation is by measuring the eigenvalues of the tidal tensor fields and assigning a cosmic web environment depending on the signature of the local tensor \citep{Hahn2007, Forero_Romero2009}. The deformation tensor is defined as the Hessian of the gravitational potential $\phi$,
\begin{equation}
    T_{ij} = -\frac{\partial^2}{\partial_{x_i} \partial_{x_j}} \phi.
\end{equation}
The eigenvalues $\lambda_0 \geq \lambda_1 \geq \lambda_2$ measure the strength of tidal compression (negative sign) or stretching (positive sign) along the principal axes. The signature of $T_{ij}$ can therefore be used to build a cosmic web map. We use the following key to classify the cosmic web environments at each volume cell:
\begin{itemize}
    \item voids: $\lambda_i < \lambda_\mathrm{threshold} \; \forall i$,
    \item walls: $\lambda_0 \geq \lambda_\mathrm{threshold} > \lambda_1 \geq \lambda_2$,
    \item filaments: $\lambda_0 \geq \lambda_1 \geq \lambda_\mathrm{threshold} > \lambda_2$,
    \item nodes: $\lambda_i \geq \lambda_\mathrm{threshold} \; \forall i$.
\end{itemize}
Similar to \citet{Forero_Romero2009}, we adopt $\lambda_\mathrm{threshold} = -0.3$, rather than $0$ to reduce noise in the classification due to small-scale structure.

\begin{figure}[tbp]
    \centering
    \includegraphics[width=\columnwidth]{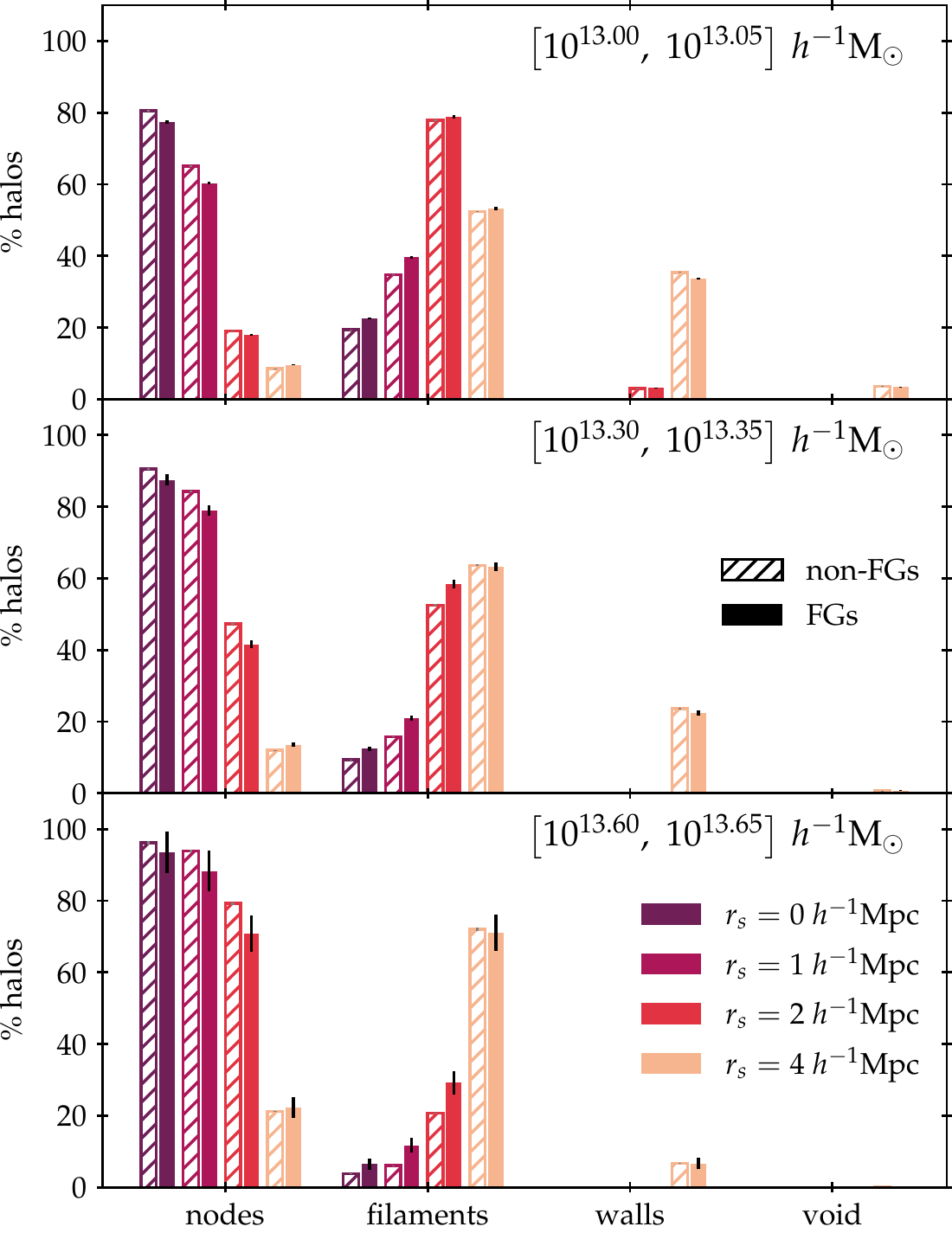}
    \caption{Percentages of halos inhabiting nodes, filaments, walls, and voids in the cosmic web. Mass bins and smoothing scales are denoted the same as in \autoref{fig:smooth_overdensities}. For each smoothing scale, the results for FG candidates are shown on the right in solid color bars, while the distribution of non-FGs are on the left in striped bars. Error bars indicate $3\sigma$ of the expected Poisson noise.}
    \label{fig:cosmic_web_signatures}
\end{figure}

\autoref{fig:cosmic_web_signatures} shows the distribution of FG candidates across cosmic web environments, in comparison to halos that do not classify as FGs. With increasing $r_s$, we remove more small-scale structure and focus on the cosmic web at larger scales. The environment to which halos are attributed therefore moves from nodes to filaments, walls, and voids at larger smoothing lengths. With group halo radii on the order of $r_{200c} \approx 0.5$\,\hMpc, values of $r_s = 1$ and $2$\,\hMpc offer a reasonable smoothing scale to blend the halo into its environment. Across all mass bins and smoothing scales, FG candidates are less likely to occupy nodes and more likely to occupy filaments or walls compared to non-FGs. This result agrees with \citet{Adami2020}, who found that fossil systems (i.e. FGs along with some poor FCs) tend to live closer to filaments than nodes. While these differences are significant by more than $3\sigma$ in all but the largest mass bin, they are small in magnitude. Thus, this slight over-representation of FG candidates in lower density cosmic web environments is unlikely to explain their unique merger history or the phenomenological differences observable at present day.

\vfil\eject\subsubsection{Clustering of Fossil Groups}

As a last environmental statistic, we measure the clustering of FGs via the two-point correlation function and compare it to the full set of halos with similar mass. For a sufficiently large random subset of the full halo sample, we expect to measure the same bias with respect to the matter auto-correlation function. Any deviation from this implies that the sample is not truly random and that there are some environmental factors that impact the probability of a halo not experiencing luminous mergers after $z=1$.

We measure the auto- and matter cross-correlation function by counting halo pairs separated by a distance $r \in [r_\mathrm{low}, r_\mathrm{high})$ for 20 linearly distributed bins between $0$ and $150$\,\hMpc. We use the \citet{Landy1993} estimator for the correlation, which has been modified for cross-correlations \citep[cf.][]{Blake2006}.
In particular, we calculate 
\begin{align}
    \xi_\mathrm{auto}(r) &= \frac{D_1 D_1 - 2 D_1 R_1 + R_1 R_1}{R_1 R_1}, \\
    \xi_\mathrm{cross}(r) &= \frac{D_1 D_2 - D_1 R_2 - D_2 R_1 + R_1 R_2}{R_1 R_2},
\end{align}
where $D_i D_j$ are the pair-counts between points in the datasets $D_i$ and $D_j$ with a separation $r$, $R_i R_j$ are the pair-counts between the randomized datasets $R_i$ and $R_j$, and $D_i R_j$ are the pair-counts between the original dataset $D_i$ and randomized set $R_j$. 
We use jackknife resampling to estimate the error of $\xi(r)$ by splitting the full simulation volume into $N_S=64$ subsets $S$ and calculating $\xi_S(r)$ by leaving out one subset each time. Then, the variance of $\xi(r)$ is given by
\begin{equation}\label{eq:jackknife_xi}
    \sigma^2_\xi(r) = \alpha(r) \frac{N_S-1}{N_S} \sum_{i=1}^{N_s} \left[\xi_s(r) - \xi(r)\right]^2,
\end{equation}
where $\alpha(r)$ is a corrective rescaling of the variance to account for the incorrect reduction of cross-subvolume pairs assumed by the jackknife technique \citep[cf. Eq. 23 in][]{Mohammad2021}.
We also calculate the matter auto-correlation function from a subset of the simulation particle distribution at $z=0$.

\begin{figure}[tbp]
    \centering
    \includegraphics[width=\columnwidth]{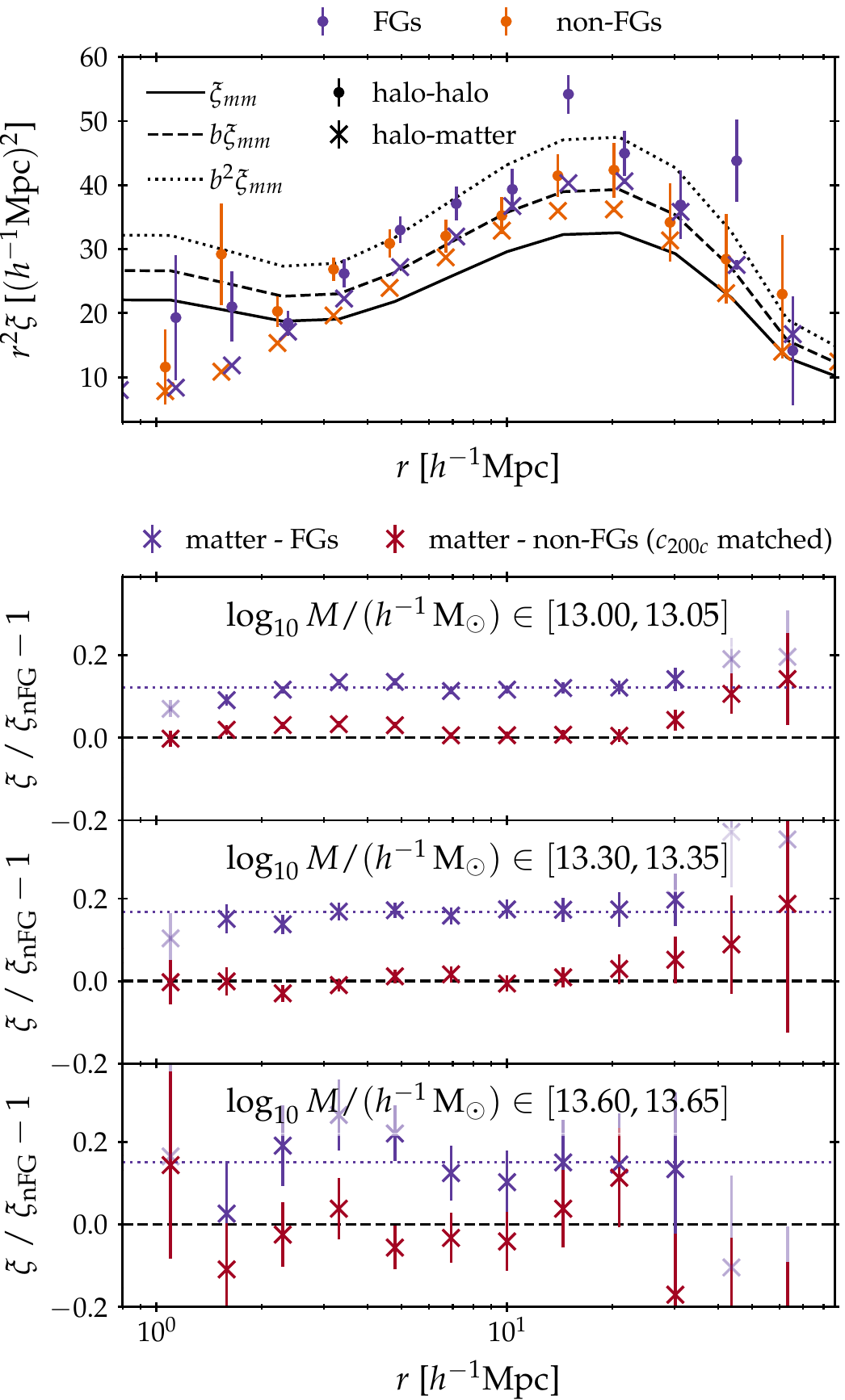}
    \caption{Comparison of the two-point auto- and matter cross-correlation function of FGs and non-FGs. 
    \textbf{Top panel:} The halo auto-correlation functions (dots) and halo-matter cross-correlation functions (crosses) for FGs (purple) and non-FGs (orange) are shown for the lowest mass bin. 
    The matter auto-correlation function is shown as the black solid line whereas the predictions of the linear bias model \citep{Cole1989} for halos of mass $10^{13}$\,\hMsun are shown in dashed and dotted lines for the cross- and auto-correlation function respectively. 
    \textbf{Lower panels:} Comparison of the fractional difference between the matter cross-correlation function of FGs (purple) with non-FGs for the three mass-bins. Additionally, we compare the full set of non-FGs and a subset that has been constrained to the concentration distribution of FGs (red; for more information, see text). Errors have been estimated by jackknife resampling of the full dataset split in 64 random groups. The horizontal dotted line indicates a $+12\%$ (top) $+17\%$ (middle), and $+15\%$ (bottom) increase we find in the halo-matter cross-correlation of FGs on scales up to 50\,\hMpc. Data points that have not been used for calculating the average increase are shown in lighter colors.
    }
    \label{fig:correlation_fct}
\end{figure}

We show our results in \autoref{fig:correlation_fct}. In the top panel, we see that on sufficiently large scales, both FGs and non-FG candidates have an increased auto-correlation and matter cross-correlation compared to the matter distribution, in agreement with expectations from the linear bias model \citep{Cole1989}. Comparing the FG bias with non-FGs at fixed mass, we find that FGs are more clustered for all mass bins. Because FG candidates have high levels of concentration (cf. \autoref{sec:dynamical_state}), we additionally compare their clustering behavior to that of non-FGs which have been constrained to the concentration distribution of the FG sample (red crosses). Examining the bias of the $c_{200c}$-matched non-FGs, we find a clustering signal compatible with the non-constrained non-FG sample (i.e. $\xi \sim \xi_\mathrm{nFG}$): this is clearly weaker than the FG clustering. Previous measurements of secondary bias \citep[e.g.][]{Wechsler2006} have found a mass-dependent relation between concentration and bias. Our results indicate that concentration by itself does not capture the underlying relation; instead, the clustering is associated with the formation time of the halo \citep[in agreement with, e.g.][]{Gao2005}. FG candidates are overwhelmingly early-forming (cf. \autoref{sec:formation_history}), whereas the concentration-matched non-FG sample contains a smaller fraction of early-forming halos, corresponding to a weaker clustering signal for this population. 

Our measurements imply that FGs are not a true random subset of all halos at fixed mass, but instead are distributed differently in the Universe. FG candidates are more clustered than non-FGs, and this behavior cannot be explained by their high concentrations, since non-FGs with high concentrations have a weaker clustering signal than do FGs. The larger FG-matter cross-correlation is qualitatively in agreement with the slightly larger density in which FGs can be found (cf. \autoref{fig:smooth_overdensities}).

\section{Summary and Discussion}\label{sec:conclusion}

In this paper, we employ Last Journey, a large gravity-only simulation, to study properties of fossil groups, based on an analysis of the merger history of halos. FGs exhibit interesting properties compared to other galaxy groups with regard to their galaxy populations and have been studied extensively in the literature (see \autoref{sec:intro}). To identify possible FG candidates in our simulation, we introduce two modeling parameters: a luminous merger mass threshold ($M_\mathrm{LM}$) and a last luminous merger redshift cut-off ($z_\mathrm{LLM}$). These parameters are motivated by the galaxy-halo connection and enable us to select a sub-group of halos that would likely host FGs. This technique does not explicitly involve a magnitude gap criterion (i.e. $\Delta m_\mathrm{12}$) and instead emphasizes the fact that FGs are deficient in L* galaxies (in keeping with the motivation of the original \citet{Jones2003} definition of FGs). We study the sensitivity of our resulting FG candidate sample with regard to the parameter choices (c.f. \autoref{fig:fraction_of_candidates}) and identify $M_\mathrm{LM} = 5\times10^{11}$\,\hMsun and $z_\mathrm{LLM}=1$ as values that result in a plausible sample when compared to observational results.

In total, we find nearly a million FG candidates in the simulation volume with masses at $z=0$ in the range $[10^{13}, 10^{14}]$\,\hMsun, corresponding to about 6\% of the total number of halos identified in this mass range (cf. \autoref{tab:fossil_groups_found}). This translates to a spatial density of $\sim 2.8 \times 10^{4}$ candidates per $(h^{-1} \mathrm{Gpc})^3$, which agrees well with predictions from previous studies \citep[e.g.][]{Jones2003, Vikhlinin1999, Santos2007, LaBarbera2009}. 

As group mass increases, the number of FGs falls sharply. For fossil clusters ($M_h \geq 10^{14}$\,\hMsun), we find only 8 candidates: less than one-thousandth of a percent of all halos in this mass bin. This result is qualitatively consistent with the literature suggesting that fossil clusters are rare; however, these estimates are a strong function of the particular fossil definition, a fact which may drive down our abundance counts compared to other definitions. Further exploration of FC definitions and corresponding abundance estimates awaits observational data from ongoing and future cluster surveys.

At the lower end of the FG mass scale, we find about 60,000 QH candidates (i.e. halos that have not had a luminous merger in their mass accretion history), most of which occupy the lowest mass bin. This constitutes $\sim6\%$ of all FG candidates. These QH candidates may represent a class of isolated giant elliptical galaxies or galaxy-poor groups.

Next, we carry out a detailed study of our large FG sample. Our investigations include analysis of FG properties such as merging probabilities, formation history, concentration and relaxation, and clustering studies. These measurements are contrasted to non-FG halos and provide clues about possible features that further distinguish FGs from non-FGs. Our major conclusions can be summarized as follows:
\begin{enumerate}[label=(\roman*)]
\item Our FG candidates experience fewer total luminous mergers in their lifetime, and the last luminous merger usually occurs earlier for FG candidates compared to non-candidates (see \autoref{fig:luminous_merging}). This is expected given our use of a last luminous merger redshift cut-off $z_\mathrm{LLM}$ to define FG candidates.
\item Looking at formation time using $z_\mathrm{frac}$ and the \diffmah parameter $\tau_c$ (which represents the transition time between fast and slow accretion phases of halo growth), we find that FG candidates accrete a significant portion of their mass earlier than do non-candidates (c.f. \autoref{fig:z_fracs} and \autoref{fig:diffmah} respectively). This result agrees with existing literature, which suggests FGs are early forming. 
\item Our \diffmah analysis further shows that FG candidates have a variety of growth rates at early ($\alpha_\mathrm{early}$) and late times ($\alpha_\mathrm{late}$) (c.f. \autoref{fig:diffmah}). Meanwhile, QH candidates are much more likely than either FG candidates or non-candidates to have both early formation times (low $\tau_c$) and high growth rates at late times (high $\alpha_\mathrm{late}$). This suggests that both FG and QH candidates have distinct mass evolution histories compared to non-candidates.
\item Our sample of FG candidates is highly relaxed and concentrated (c.f. \autoref{fig:contours} and \autoref{fig:cdeltas_corrected}). The objects also contain a  significantly smaller fraction of substructure -- both in terms of the largest subhalo and of the sum total of all subhalos --  than non-candidates (c.f. \autoref{fig:fsub_stats} and \autoref{tab:fsub_stats}). All these results indicate that our FG candidates are in a state of high dynamical relaxation, in agreement with existing literature.
\item Looking at local overdensities and the cosmic web environment, we do not find any strong evidence of environmental effects on the probability of luminous merging as it relates to the formation of FGs (c.f. \autoref{fig:smooth_overdensities} and \autoref{fig:cosmic_web_signatures} respectively). The existing literature has not come to any collective agreement about the role of environment in the formation of FGs, however, our results do support the conclusions of particular studies \citep[e.g.][]{vonBendaBeckmann2008}.
\item Our sample of FG candidates is more clustered than non-FGs in the same mass range (c.f. \autoref{fig:correlation_fct}). This pattern is not explained by higher concentration among FG candidates compared to non-FG candidates, as the excess clustering signal disappears when we perform concentration matching on the non-FG sample. This implies that our luminous merger criterion -- which is highly correlated with early halo formation time -- is a strong indicator of assembly bias.
\end{enumerate}

In summary, our population of FG candidates is characterized by fewer/earlier mergers, unique patterns of evolutionary history, high dynamical relaxation, little correlation with environment, and a noticeable clustering signal compared to non-FG candidates. This set of features is consistent with those observed in spectroscopically confirmed FGs. It is significant that our choice of last luminous merger criterion and redshift cutoff have led to a well-matched candidate population from merger trees in a gravity-only simulation that does not account for any gas physics.

Although our FG candidates display many unique features, their extreme behavior does not indicate that they are fundamentally distinct from other more generic halos. Rather, since halo merging is a discrete phenomenon, we expect to find low values in the distribution of merging activity: FGs appear to occupy these regions. They do not form a separate, distinct peak in an otherwise continuous distribution.
The large volume of Last Journey allows us to view halo merging distributions and their tails with significantly improved statistics, thus robustly identifying a large number of relatively rare objects such as FGs. We therefore concur with \citet{Aguerri2021}'s interpretation that FGs are simply an extreme population produced according to the normal structure formation processes.

In the future, we will extend our studies to include more realistic information about the galaxy populations in halos that can be obtained via semi-analytic modeling and hydrodynamical simulations with full baryonic physics. The ever-growing power of supercomputers will enable studies of similar sample size as we used in this paper on simulations that include accurate modeling of galaxy formation. 
Additionally, upcoming surveys such as the Legacy Survey of Space and Time \citep[LSST,][]{LSST2019} conducted at the Vera C. Rubin Observatory will provide larger observational samples that will enable more refined studies of FGs and provide more input into the modeling assumptions as well. 
The improved statistics from both observations and simulation approaches will provide a better understanding of the role FGs play in our Universe.

\begin{acknowledgments}
We thank Lindsey Bleem, Patricia Larsen, and Ann Zabludoff for helpful discussions and interesting suggestions, as well as Andrew Hearin and Alex Alarcon for useful insights into the results of our \diffmah-based analysis.

Argonne National Laboratory's work was supported under the
U.S. Department of Energy contract DE-AC02-06CH11357.  An award of computer time was provided by the ASCR Leadership Computing Challenge (ALCC) program. This research was supported in part by DOE HEP's Computational HEP program. This research used resources of the Argonne Leadership Computing Facility at the Argonne National Laboratory, which is supported by the Office
of Science of the U.S. Department of Energy under Contract
No. DE-AC02-06CH11357. We are indebted to the ALCF team for their outstanding support and help in enabling us to carry out a full-scale simulation on Mira. 
\end{acknowledgments}

\bibliography{references}
\bibliographystyle{aasjournal}

\end{document}